\documentclass[sigconf,nonacm]{acmart}
\usepackage{booktabs}
\usepackage{graphicx}
\usepackage{placeins}
\usepackage{stfloats}

\makeatletter
\g@addto@macro\UrlBreaks{\UrlOrds}
\makeatother
\renewcommand\footnotetextcopyrightpermission[1]{}
\begin{document}

\title[Plausible Patients, Impossible Populations]{Plausible Patients, Impossible Populations: Auditing Epidemiological Fidelity in Large Language Model Mental Health Simulations}
\author{Patrick Keough}
\affiliation{%
  \institution{Independent Researcher}
  \city{Sydney}
  \country{Australia}}
\email{pskeough@gmail.com}

\begin{abstract}
Language models asked to simulate psychiatric patients produce cases that survive inspection one at a time and populations that match no real one. We gave GPT-4o-mini, Gemini-3-Flash, DeepSeek-V3 and GLM-4.7 each of 120 demographic cohorts under two framings, one written as a clinician enters a patient and one as a person describes themselves, and scored all 28,800 responses against survey-weighted PHQ-8 anchors derived from NHANES microdata. Case by case the output holds up: 97.3\% of elevated presentations satisfy the DSM-5 gateway rule, violating it at 2.68\% against a chance null of 10.4\%. As populations, four things fail at once. Every benchmarkable group returns inflated by 2.8 to 5.5 PHQ-8 points, and 18.2\% of simulated patients screen at the treatment threshold against 7.5\% of adults. Population Black-White and Hispanic-White disparities do not survive the simulation, with two models attenuating each gap and two flattening or inverting it. Symptom covariance reorganizes by cohort, putting the error beyond the reach of any recalibration, and demographic offsets do not stack, so a correction fitted on marginals misses the cells by about 0.4 points in either direction. And the answer does not hold still: at the decoding a deployment inherits, a third of patients change severity category between two draws of one prompt and one in five crosses the line separating watchful waiting from treatment. We read the four together as one failure, and name the gap between case-level plausibility and population-level failure the coherence-fidelity dissociation. Individual cases clear a formal rule a case review would apply; the populations they compose fail every comparison we can construct against a real one. Gender identity carries an extreme on both symptom structure and regeneration stability, and no federal benchmark exists against which any of it can be checked. The patients look right. They do not represent real populations.
\end{abstract}

\maketitle

\section{Introduction}

Large language models are being deployed in mental health work faster than they are being evaluated. We ask whether these models resemble the epidemiology of a real population, and whether evaluation at scale reveals latent biases in how they represent real demographics. Individual outputs present convincing profiles, and the models rarely produce a clinically impossible patient (Section~\ref{sec:coherence}), which leaves both clinical and personal users liable to accept the model's projection as accurate. How accurate those projections are at scale has not been evaluated, and we set out to evaluate it.

These models are being used in mental health care, where a wrong answer carries consequences for health outcomes, and what they encode about race, gender identity, income and partnership has not been measured against the populations they stand in for. That creates two problems at once: individuals potentially misdiagnosed, and systemic inequities compounding at scale. Clinicians using these tools might believe or follow a biased result before patient consultation, and individual users seeking help might be systematically misled the same way. No single encounter shows what those errors add up to across a population.

Whatever bias these models carry is encoded internally, in weights nobody can inspect. It has to be characterized from the outside, by holding a large enough grid of outputs against a population whose values are known. Algorithmic fairness has established bias this way before, and what that work found was a difference in error rates rather than in average error (Section~\ref{sec:related}). We apply the same logic across a demographic grid.

We ask what these models produce when asked for a population rather than a patient: whether the distributions they generate match the ones they are standing in for, whether the error is uniform across demographic groups, and where it concentrates.

We took four LLMs from the cost tier a high-volume simulation pipeline would actually run on: GPT-4o-mini, Gemini-3-Flash, DeepSeek-V3 and GLM-4.7. We ran them on a psychological battery including the PHQ-8~\cite{kroenke2009phq8}, the GAD-7~\cite{spitzer2006gad7} and the AUDIT-C~\cite{bush1998auditc}, giving each one of 120 group identities through two framing cases. They were chosen on deployment likelihood rather than benchmark standing: two long-available cheap endpoints a pipeline built in the preceding period would still be calling, and two a pipeline built at collection time would reach for first. Each model ran 30 times on each prompt and condition, to control for response variance and to check internal stability, for 28,800 responses in total. We then ran a series of statistical tests to derive their proximity to real population data taken from NHANES. That analysis examined individual group effects, framing variance, and differences between the models themselves. Our aim was to use AI responses at scale to better understand how these systems represent bias across different population groups, and whether the framing an agent is given changes their response patterns.

Likert-based psychometric batteries give us a quantity we can check: every demographic group has an expected score in the population, and the model's answers fall on the same scale. With eight ordinal symptom items on the PHQ-8, a cohort of responses yields a covariance matrix and a dispersion. Examining how a model scores a given group across all 120 cohorts lets us find cross-sectional patterns of representation for specific subgroups, and derive the model's accuracy in both individual and cross-sectional terms. We run the full factorial over race, gender identity, socioeconomic status, and relationship status. Bias then becomes a pattern across those axes, which gives us insight into what these models hold. We went in expecting bias to take roughly the same form on each axis, differing in size. Section~\ref{sec:structure} shows it does not.

Across four models and 28,800 simulated responses, these models do not resemble the epidemiology of any real population. Every demographic group we can benchmark against nationally representative data comes back inflated by 2.8 to 5.5 PHQ-8 points when scored against survey-weighted NHANES anchors. The gradient is steepest at the bottom of the income range, where low-income personas run 5.48 points above the population against 4.24 for the next largest group. Conditioning that one anchor on the enrolment the persona text states moves it to 4.21 and the range to 2.8 to 4.2, at which point the design no longer separates it from the others. Alongside that, only 2.68\% of elevated cases violate DSM-5 gateway requirements, against a chance null of 10.4\%. Formal single-case validation cannot see the failure at all, and we call that the coherence-fidelity dissociation.

We contribute four results. Overestimation is systematic, appearing in every group we can benchmark and surviving every population comparator we can construct, including a bound built on clinical anchors. One anchor reverses it: a treatment-entry cohort selected on high severity. Appendix~\ref{app:anchors} works through why. The error is differential rather than a uniform offset, and on two of three racial contrasts the disparity the population carries does not survive into the simulation. No correction we can think of reaches it, since the bias survives both framings, does not hold still under regeneration, reorganizes symptom structure, and does not stack additively across the axes that produce it. And the models disagree about how to apply demographic conditioning rather than about whether to: demographic factors alone carry 70.3 percentage points of cell-mean variance against 1.5 for model identity, while the model-by-demographic terms carry roughly twenty times per term what the demographic interactions do.

\section{Related Work}
\label{sec:related}

\textbf{Bias in clinical AI.} Algorithmic bias in health care is well documented, from racial bias in care-management risk scores~\cite{obermeyer2019dissecting} to race-based errors propagated by LLMs in medical question answering~\cite{omiye2023racebased}, and recent evaluations extend this to generative settings, where GPT-4 reproduces demographic disparities in diagnosis and treatment recommendation~\cite{zack2024gpt4}, and where rubric-based human evaluation surfaces equity-related harms in long-form medical answers~\cite{pfohl2024toolbox}. The racial bias in the COMPAS recidivism scores was established without any access to the algorithm at all, by comparing its outputs against real outcomes, which showed Black defendants disproportionately misclassified as high risk and white defendants as low~\cite{angwin2016machinebias,dressel2018accuracy}. What that work found was a difference in error rates, not in average error. Obermeyer et al.\ found their bias the other way, by asking what the algorithm was actually predicting, which turned out to be cost, not illness. We cannot ask that question of a language model. We ask the COMPAS question instead: hold the outputs against a real population and see where the error concentrates.

\textbf{Simulating respondents.} Argyle et al.~\cite{argyle2023outofone} introduced silicon sampling, reporting that properly conditioned models emulate response distributions across human subgroups. Later work qualified that in three ways this audit reproduces on a clinical instrument. Bisbee et al.~\cite{bisbee2024synthetic} found less variation in synthetic survey responses than in the surveys they replace; Santurkar et al.~\cite{santurkar2023whose} found model opinion distributions misaligned with all sixty US demographic groups examined, persisting after explicit steering; Cheng et al.~\cite{cheng2023compost} characterized caricature as individuation plus exaggeration and found simulations of marginalized groups most susceptible. Our severity residuals put a number on what Bisbee et al.\ reported qualitatively. We do not offer dispersion ratios as the analogous measurement, for the reason Section~\ref{sec:estimand} gives, and the covariance reorganization of Section~\ref{sec:covariance} carries the caricature construct from open-ended text into psychometric structure, an operationalization that work did not test.

\textbf{Where the dissociation sits.} Realism at the level of the individual sample alongside failure at the level of the distribution is the precision-recall decomposition for generative models~\cite{sajjadi2018precision,kynkaanniemi2019improved}, and the shape we find, no severe cases in 14,400 draws with mass concentrated in one five-point band, is the low-recall side of it. We add a case-level criterion with clinical content rather than a density estimate: a formal DSM-5 gateway rule, referred to three constructed nulls, on an instrument whose scoring is defined outside the model. The dissociation we name is that criterion instantiated, not a new phenomenon in generative modelling.

\textbf{Two results that complicate ours.} Villarreal-Zegarra and Bellido-Boza~\cite{villarrealzegarra2026synthetic} fitted confirmatory factor models to synthetic PHQ-9, PHQ-8 and GAD-7 responses and reported good fit and high internal consistency. We run a different test, since separating the two results requires fitting the same factor model \emph{within} each demographic cohort, which Section~\ref{sec:covariance} does and finds model-dependent. Second, Cheng et al.~\cite{cheng2023marked} document strong racial stereotyping in persona-conditioned text, while racial contrasts are the smallest structural effects in our design. Both can hold if racial bias in these models lives in lexical and narrative content and not in the covariance of instrument items, but we report it as an open question.

\section{Method}

\subsection{Design}
\label{sec:design}

Four models each completed 7,200 structured psychiatric assessments: GPT-4o-mini, Gemini-3-Flash, DeepSeek-V3 and GLM-4.7, routed through OpenRouter. Two of the four routes resolve to dated snapshots, \texttt{z-ai/glm-4.7-20251222} and \texttt{google/gemini-3-flash-preview-20251217}, and Appendix~\ref{app:prompts} gives every endpoint string, its resolved slug, and the upstream provider that served each call. The panel sits at the small-to-mid cost tier that a high-volume simulation pipeline would actually run, and we fixed composition at two US-developed and two China-developed models, a balanced sampling on development context rather than a hypothesis about it. Every model ran at its provider's defaults, under which GLM-4.7 alone emitted reasoning tokens. Any cross-model comparison carries that difference alongside model identity. Claims here describe this cost tier as it stood in early 2026.

Selection followed deployment likelihood at the time of collection, not benchmark standing. Gemini-3-Flash had just become the default model behind AI Mode in Google Search~\cite{google2025aimode}, making it the newest cheap-tier model from the largest-distribution vendor, though that is consumer query volume and not the API workload this panel stands in for. GLM-4.7 was the current open-weight flagship at cost-tier pricing, the position a cost-sensitive builder selects from. GPT-4o-mini and DeepSeek-V3 enter for the opposite reason, as long-available cheap endpoints of the preceding period rather than the current frontier. The panel therefore spans two release generations, the dated snapshots on one side and the established routes on the other, and covers both what a pipeline would adopt at collection time and what one built earlier would still be running.

The design crosses 120 demographic cohorts, race (5) $\times$ gender identity (4) $\times$ socioeconomic status (3) $\times$ relationship status (2), with 30 iterations per cohort under each of two prompt conditions, for 28,800 simulations. The system prompt asked for accuracy, not performance. It instructed each model to base its responses on the statistical likelihood of symptoms for the demographic intersection in front of it, referencing federal population sources by name, NSDUH and CDC, and to return integers only. NSDUH is the looser of the two. It assesses depression against DSM criteria rather than administering the PHQ-8, and cannot produce the totals we benchmark. Appendix~\ref{app:prompts} reproduces the system prompt in full. Each model was given the four instruments by name, with their item counts and response scales, and returned a single JSON object of integer item scores covering all four, with the item wording never shown. We therefore measure each model's own representation of what the instruments ask, rather than its ability to score a form we supplied. The model chooses which symptom occupies which position, and the scoring assumes that choice is canonical. Asked to enumerate the eight PHQ-8 items in the order it would score them, every model returns \texttt{DPQ010} through \texttt{DPQ080} in the standard sequence on 40 of 40 attempts, including the two positions the gateway rule of Section~\ref{sec:coherence} reads. The probe postdates collection and is reported in full in Appendix~\ref{app:prompts}. The PHQ-8 omits the ninth PHQ-9 item to avoid triggering API safety refusals.

Every cell is populated, though the four models did not arrive there on equal terms. In the narrative run GPT-4o-mini and Gemini-3-Flash each filled all 3,600 of their cells in 3,600 calls. DeepSeek-V3 needed 3,935. GLM-4.7 needed 6,141, returning an error rather than a response on 21\% of them. In the clinical run 62 GLM-4.7 rows were completed by hand instead of through the API, and Appendix~\ref{app:prompts} records that recovery in full alongside the call ledger. Section~\ref{sec:modelfactor} returns to what model identity costs a pipeline.

The two conditions are not a prompt-engineering manipulation. They are the two ways this reaches a person. The clinical condition presents the demographic profile in the third person as structured intake data, the way a clinician would enter a patient into a tool. The narrative condition presents the same profile as a first-person self-description, the way the person themselves would ask. Appendix~\ref{app:prompts} reproduces both templates verbatim, alongside the deterministic map that generates the second from the first.

\subsection{Estimand and decoding}
\label{sec:estimand}

Each simulation conditions a model on a single persona vignette and elicits one assessment. The quantity measured is the model's conditional response distribution given a demographic cell, not a sample of distinct individuals. Comparisons of means against population means are well posed under this design. Comparisons of variance are not: a single fixed vignette is never asked to produce the between-person heterogeneity that carries most of the variance in a real PHQ-8 distribution. No variance-fidelity claim follows, and every dispersion result below compares cohorts against each other inside one model, never against a population variance this design cannot supply.

Composition follows from the same discipline. The factorial grid is uniform. Half of the simulations behind every race or SES marginal come from transgender personas, while the population anchor for that marginal reflects actual composition, in which transgender adults number under one percent and cannot be identified at all. Pooled marginals accordingly run 1.2 to 1.8 points above their cisgender-only counterparts. We compute benchmarked residuals on cisgender personas, matching the comparison frame to the anchor, and retain full-grid pooled residuals as a sensitivity column. Direction and significance of every finding survive either choice.

The results divide on whether they need the population anchor. Level residuals need it and inherit every objection it carries, administration mode among them (Section~\ref{sec:limits}). Between-group contrasts are free of it. Any anchor-side effect common to both groups differences out, leaving a comparison that survives a mis-specified anchor. We rest the differential findings on that property. It requires the anchor-side effect to be additive and shared, and three things would break it: a mode or frame effect that scales with severity rather than shifting it, differential item functioning across the groups being compared, and any anchor error that itself differs by group. Section~\ref{sec:limits} takes each in turn.

Decoding was fixed by the same discipline. Every model ran at its provider's default sampling, the configuration a deployment inherits when nothing is set. A pre-registered control repeated twelve cohorts at temperature 0, 2,880 generations in all (Appendix~\ref{app:decoding}). The remedy does not port across endpoints. Bootstrapping the twelve cohorts, category flipping falls 35.8 points on GPT-4o-mini (95\% CI $-41.8$ to $-29.1$) and by an amount not distinguishable from zero on GLM-4.7 ($-3.7$, $-9.8$ to $+2.0$). Where it falls furthest the generation has stopped varying, returning one repeated total in 58 to 75 percent of cells. We read that as a sampler being silenced and not as a judgement steadying. Flip rates below are reported at the decoding a deployment inherits.

\subsection{Ground truth: derivation and validation}
\label{sec:groundtruth}

All depression benchmarks are computed by the author from NHANES public microdata, cycles 2005--2006 through 2017--2018, merged on respondent ID. PHQ-8 totals sum the eight depression-questionnaire items \texttt{DPQ010} through \texttt{DPQ080}, which are the interest, mood, sleep, energy, appetite, self-worth, concentration and psychomotor items, for complete responders aged 18 and over. Estimates use MEC examination weights. Means and standard deviations are nationally representative population parameters on the same 0--24 scale the models produce. Race groups use the recoded race and ethnicity variable \texttt{RIDRETH1}, with Asian identified from the expanded \texttt{RIDRETH3} for the 2011+ cycles that carry it. The Asian anchor rests on a shorter window than the others. Income groups use the poverty-income ratio at two federal policy cutpoints.

The pipeline is validated against two published targets before being trusted for new numbers. Restricted to 2005--2016 and unweighted it reproduces the PHQ-9 descriptive table of Patel et al.~\cite{patel2019phq9} to within 0.02 on every mean and standard deviation but one, and restricted to 2013--2016 and weighted it reproduces the national prevalence estimates of Brody et al.~\cite{brody2018depression} exactly (8.1\% overall, 10.4\% women, 5.5\% men). Three cells have no defensible benchmark and are reported without residuals: transgender women, transgender men, and multiracial adults.

One threat to the anchor deserves its own treatment. Population depression rose over the anchor window and rose again after 2020, while the models were queried in 2025. An anchor pooling 2005--2018 could be benchmarking against a population that no longer exists. We test this by rebuilding every anchor twice more under the same derivation and frame, on the 2017 to March 2020 pre-pandemic file and on the 2021--2023 cycle, each with its own weights and strata. The all-adult weighted mean runs 2.98 on the paper window, 3.14 pre-pandemic and 3.97 in 2021--2023. The 2021--2023 file also changes how the screener was administered, from the interviewer-administered MEC questionnaire used through 2020 to audio computer-assisted self-interview. Self-administration raises endorsement on stigmatised items, so part of that rise is mode rather than population. Every residual shrinks by 0.86 to 1.28 points against the most recent cycle and none closes; the smallest falls from +2.75 to +1.83. The recent anchors are noisier, one cycle replacing seven, so the full-window anchors stay primary.

\subsection{Statistical reporting}
\label{sec:stats}

Every statistic is computed on cell means rather than on raw generations. The unit of analysis is the design cell, a model crossed with a demographic cohort, of which there are 480 per condition. The 30 iterations inside a cell are repeated draws from one conditional response distribution, not distinguishable individuals. Treating the 28,800 generations as exchangeable would understate standard errors four to seven fold. Cells within a model are not independent either, and two-way cluster-robust errors over model and cohort run 1.8 to 4.7 times the between-cell errors. Across the residual tests of Table~\ref{tab:residuals} that changes no sign or significance. It was not applied to the racial contrasts of Section~\ref{sec:crossgroup}, which are paired within model on 96 pairs and carry four clusters at the model level, too few for a cluster-robust interval to be trusted. Those contrasts are reported with their per-model estimates for that reason, and every interval in this paper is within-model precision rather than an interval on the model class.

The cohorts under test are also the cohorts with elevated means. Covariance and dispersion comparisons are referred to nulls that hold severity fixed. NHANES anchors carry sampling error of their own, computed by Taylor linearisation and small enough to treat as fixed: the largest is 0.149 points on the persona-matched low-income anchor, whose smaller subpopulation costs precision, and 0.078 across the demographic anchors of Table~\ref{tab:residuals}, against a smallest reported residual of +2.75. Residuals divided by the population standard deviation of the corresponding group are written $d_{\mathrm{pop}}$ and read as a common scale and not as Cohen's $d$. Appendix~\ref{app:inference} gives the constructions, the tolerances, the cluster bootstrap, and the reasons the simpler controls were rejected. Inference is organized into four pre-specified families, each Benjamini-Hochberg adjusted within itself: residuals and design contrasts (34 tests), racial contrasts against zero and against the population (3 and 3), the severity-matched covariance permutations of Section~\ref{sec:covariance}, and the dispersion matching tests. We do not correct across families and report the structure so a reader can. The 34-test family ships as a machine-readable ledger and is summarized in Appendix~\ref{app:ledger}.

\section{Results}

\subsection{Severity is inflated in every benchmarkable group}
\label{sec:overestimation}

Table~\ref{tab:residuals} reports the central result. Every demographic group with a valid benchmark receives simulated PHQ-8 severity far above its population value, in all four models pooled and in all 36 model-by-group cells individually.

\begin{table}
\caption{Severity residuals against ground truth (model mean minus NHANES weighted mean), cisgender personas; pooled full-grid residual as sensitivity. Intervals are clustered at the design cell (Section~\ref{sec:stats}). The persona-matched row conditions the low-income anchor on both income and Medicaid enrolment to match the persona text (Appendix~\ref{app:anchors}); every other row conditions on demographics alone. $\dagger$ its bracket covers anchor sampling error alone, design SE 0.149 on $n = 2{,}736$, where the other rows carry design-cell clustering on the simulation side.}
\label{tab:residuals}
\small
\setlength{\tabcolsep}{3pt}
\begin{tabular}{lcccc}
\toprule
Group & GT mean (SD) & Residual [95\% CI] & $d_{\mathrm{pop}}$ & Pooled \\
\midrule
White & 2.91 (3.83) & +4.24 [+3.43, +5.04] & 1.10 & +5.77 \\
Black & 3.24 (4.25) & +3.77 [+3.03, +4.52] & 0.89 & +5.27 \\
Asian & 2.16 (3.01) & +3.92 [+3.33, +4.51] & 1.30 & +5.55 \\
Hispanic & 3.15 (4.15) & +3.96 [+3.20, +4.72] & 0.95 & +5.49 \\
Cis men & 2.50 (3.58) & +4.07 [+3.60, +4.53] & 1.13 & n/a \\
Cis women & 3.44 (4.19) & +3.97 [+3.49, +4.45] & 0.95 & n/a \\
Low SES & 4.25 (4.89) & +5.48 [+5.12, +5.83] & 1.12 & +6.99 \\
\quad persona-matched & 5.52 (5.60) & +4.21 [+3.92, +4.50]\textsuperscript{$\dagger$} & 0.75 & \\
Middle SES & 3.12 (3.95) & +3.15 [+2.73, +3.57] & 0.80 & +4.96 \\
High SES & 2.23 (3.08) & +2.75 [+2.38, +3.12] & 0.89 & +3.99 \\
\bottomrule
\end{tabular}
\end{table}

\begin{figure}[b]
\includegraphics[width=0.97\linewidth]{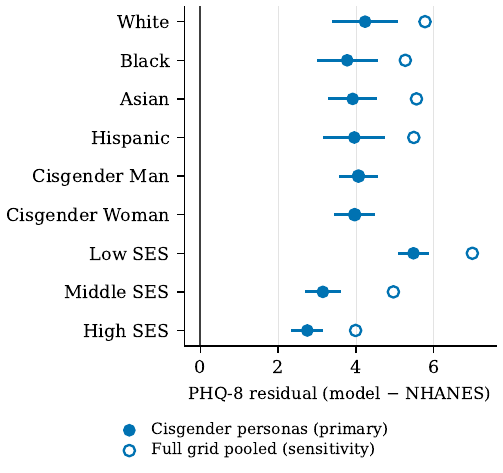}
\caption{PHQ-8 severity residuals against NHANES weighted anchors. Filled markers: cisgender personas (primary frame, 95\% CIs). Open markers: full persona grid pooled (sensitivity).}
\label{fig:residuals}
\end{figure}

The pattern is uniform in direction and graded in magnitude, every $d_{\mathrm{pop}}$ at or above 0.80 and every interval excluding zero by a wide margin (Figure~\ref{fig:residuals}). The socioeconomic gradient is preserved in direction and inflated in size, population low minus high being 2.02 points against a simulated 4.75. That inflation is not a separate effect at the endpoints: least squares through the origin puts the multiplier at 2.29 and the log-scale estimator at 2.32, either of which predicts a simulated gradient of 4.62 to 4.69 against the observed 4.75. Inside the endpoints no constant reproduces it. The same multipliers under-predict low-to-middle (2.58 against an observed 3.46) and over-predict middle-to-high (2.04 against 1.29), each by about a third and in opposite directions, and the multiplier itself runs from 2.01 for middle-income personas to 2.82 for Asian personas. Measured against its own population distribution the largest miscalibration in the design belongs to Asian personas, not to low-income ones. In practical terms the simulated low-income patient sits at the top of the mild band, within a point of the moderate threshold, while the typical real low-income adult scores in the minimal band (weighted median 3, with 65.5\% at 4 or below).

Restated as the quantity a service would actually plan around, 18.16\% of simulated cisgender generations screen at PHQ-8 = 10 or above against 7.48\% of the population on the same anchor, a factor of 2.43, with a per-model range running from 9.36\% for DeepSeek-V3 to 23.61\% for Gemini-3-Flash. A screening cohort drawn from these models carries roughly two and a half times the caseload the population would generate. That restatement inherits the administration-mode bound of Section~\ref{sec:limits} along with the residual it restates. For scale, the minimum important difference for individual change on the nine-item form is 5 points~\cite{lowe2004monitoring}. These residuals reach the same order of magnitude. They are group means on the eight-item form, and the two quantities do not admit direct comparison.

These models get the shape of the distribution as wrong as the level. The population distribution is heavily right-skewed, with 77.1\% of adults at 4 or below and a long thin tail, while the pooled simulated draws put 55.3\% into the single five-point band from 5 to 9 against 15.4\% in the population (Appendix~\ref{app:shape}). Severe scores of 20 or above are 0.6\% of the population, about one adult in 170, and not one of the 14,400 cisgender generations reached that range. Across the full grid sixteen generations do, every one of them a transgender persona. The interior of the scale fills the space instead, on every instrument the models answer. The top category takes 0.92\% of PHQ-8 item responses, 0.52\% of GAD-7 and 0.17\% of AUDIT-C against 3.93\% in the population's PHQ-8 items, the bottom category takes 33.5\%, 31.4\% and 24.4\% against a population 74.8\%, and 66 to 75\% of simulated item responses fall strictly between the two endpoints against 21\% in the population. Systematic, content-independent response artifacts on survey instruments are a documented property of language models~\cite{dominguezolmedo2024questioning}. The artifact here is endpoint avoidance, and it holds across three instruments at once inside the same generations. A pipeline drawing cisgender patients from these models therefore receives a corpus with no severe cases in it, and from two of the four models almost no floor mass either: DeepSeek-V3 puts 3.7\% of draws at 4 or below and GPT-4o-mini 4.8\%, against 50.7\% for Gemini-3-Flash and 47.1\% for GLM-4.7 (Appendix~\ref{app:shape}). Any procedure trained or evaluated on such a corpus has never seen the patients it exists to catch.

The choice of comparator can be varied, and one variation raises the residual. If what the prompt elicits is a modal patient and not a mean one, the mode is the right comparator. The simulated cisgender distribution has a mode of 8 and a median of 7; the population's are 0 and 2. Read against the population median the pooled residual is +4.99 and against the mode +6.99, both larger than the +4.01 we report against the mean. We report the mean comparison throughout, as the most conservative of the three. The prompt also frames a screening encounter, where people score higher than an unselected household sample. Appendix~\ref{app:anchors} bounds that against published clinical anchors and finds roughly half of the pooled residual and half to two-thirds of the low-income one surviving a comparison against an unconditioned primary-care caseload. The low-income persona names Medicaid as well as income; the anchor conditions on the poverty-income ratio alone. Among adults earning under \$35,000, those on Medicaid score 1.82 points above those who are not. Conditioning the anchor on both, to match the persona text, moves the low-income benchmark from 4.25 to 5.52 and the residual from $+5.48$ to $+4.21$ (Appendix~\ref{app:anchors}).

\subsection{The error is not uniform, and on race the disparities flatten}
\label{sec:crossgroup}

Uniform overestimation would be a calibration problem. The residuals in Table~\ref{tab:residuals} are not uniform.

Models can be miscalibrated in level while still reproducing the shape of population differences, and our ground truth measures one cross-group gap cleanly enough to test that: White minus Asian, a population value of 0.755 points. No model reproduces it, and the errors follow no pattern across models. DeepSeek-V3 comes closest at a simulated gap of 0.708, GLM-4.7 is furthest at 1.736, and the two US-developed models bracket the target from either side, GPT-4o-mini undershooting at 0.403 and Gemini-3-Flash overshooting at 1.447. One China-developed model sits at each extreme. Development context organizes nothing here.

We had expected the racial contrasts to be attenuated, because a model with a weak grip on a small population difference should attenuate it. On two of the three, attenuation is complete. We test each contrast against the population value it should reproduce, pairing cisgender cells within model, condition, gender, socioeconomic level and relationship status so that the two cells in a pair differ in race alone, which gives 96 matched pairs per contrast (Appendix~\ref{app:inference}).

The population places Black adults 0.330 points above White adults. These models place them 0.131 \emph{below}, on an interval of $[-0.273, +0.011]$ covering zero ($q = 0.11$) and sitting 0.461 points under the population value ($q < 0.001$). We computed two intervals and report the conservative one: a bootstrap holding the panel composition fixed puts this contrast below zero, and Appendix~\ref{app:inference} explains our choice of the wider one. Hispanic against White runs the same way, a population gap of $+0.243$ against a simulated $-0.034$, again indistinguishable from zero ($q = 0.57$) and again 0.277 points under the population ($q = 0.009$). We cannot say these models reverse those two disparities, only that the disparities do not survive into the simulation. Both contrasts are equivalent to zero within a bound set at the population disparity itself, at $p = .004$ for Black against White and $p = .0003$ for Hispanic against White by two one-sided tests on the same 96 matched pairs. The design would have detected a gap the size of the population's with probability $.995$. These are bounded nulls rather than underpowered ones, and the bounds are wide. Equivalence holds for the Black contrast only at 0.251 points or wider and fails at half the population disparity, and its 90\% interval $[-0.251, -0.010]$ sits wholly below zero. The simulated gap is bounded well under the population value and runs the other way. The groups do not come back identical. The Asian contrast is the one they keep, at $-1.074$ against a population $-0.755$, exaggerated by about two fifths, an exaggeration that does not survive correction ($q = 0.07$). That figure carries a caveat the other contrasts do not. The Asian anchor rests on 2011 and later cycles while the White anchor pools 2005--2018, and matching the windows moves the population gap from 0.755 to 0.840 and the exaggeration from 42\% to 28\%. The driver is a single anomalously low 2005--2006 White cycle rather than any secular trend, and Appendix~\ref{app:anchors} reports the matched-window figures.

\begin{figure}
\includegraphics[width=0.97\linewidth]{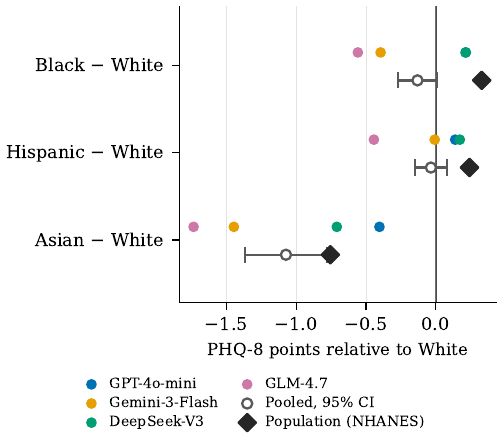}
\caption{Each racial group's simulated severity relative to White personas, against the gap the population actually shows. The pooled estimate carries its 95\% interval over 96 matched cell pairs. On Black and Hispanic that interval covers zero while the population marker sits well outside it. The disparity the population carries does not survive into the simulation. The Asian contrast keeps its direction and is exaggerated. Per-model points show the spread behind each pooled value.}
\label{fig:crossgroup}
\end{figure}

Pooling hides a split worth reporting. The negative estimates on Black minus White belong to two of the four models, Gemini-3-Flash at $-0.394$ and GLM-4.7 at $-0.558$, while DeepSeek-V3 and GPT-4o-mini sit at $+0.214$ and $+0.217$, on the same side of zero as the population (Figure~\ref{fig:crossgroup}). None of the four reproduces the population gap and they miss it in both directions, the pattern Section~\ref{sec:modelfactor} finds across the whole design.

A flattened disparity is a different kind of error from an inflated level. A scaling problem leaves the between-group structure intact underneath it, and these models do not leave it intact. A synthetic cohort drawn from them carries the inflation and almost none of the racial structure a disparities analysis would be looking for. A study run on that cohort would recover a gap far under the one the population carries, and on the Hispanic contrast would report a null.

Relationship status behaves the same way and cannot be benchmarked. Simulated single personas score 0.88 points above married personas pooled, and the gap is again model-specific, running +1.57 in GLM-4.7 and +1.30 in Gemini-3-Flash against +0.42 in DeepSeek-V3 and +0.23 in GPT-4o-mini. The direction is epidemiologically plausible, partnered status being broadly protective, but no representative severity norm stratified by relationship status exists and we make no fidelity claim about it.

\subsection{It survives both ways this reaches a person}
\label{sec:framing}

The clinical and narrative conditions are the two sides of a real encounter. The contrast between them measures what a user would experience and not what a prompt engineer would tune. Pooled across models, the clinician-side framing raises PHQ-8 severity by 0.32 points (95\% CI 0.24 to 0.41, paired $t = 7.29$ over 480 design cells, $d_z = 0.33$), with the same pattern on GAD-7 and a much smaller shift on AUDIT-C.

\begin{figure}
\includegraphics[width=\linewidth]{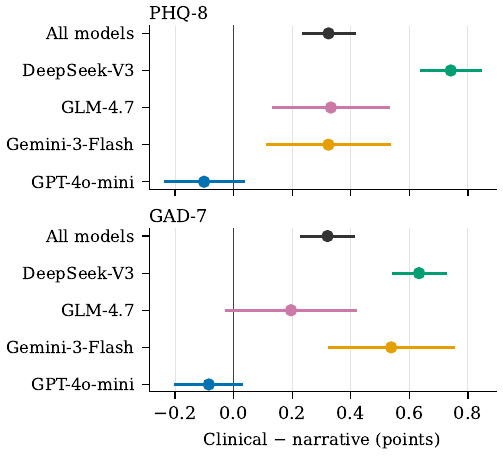}
\caption{Condition contrast (clinical minus narrative) with 95\% CIs, pooled and per model, on PHQ-8 and GAD-7. The effect is model-specific and undetectable in GPT-4o-mini, whose interval spans zero.}
\label{fig:condition}
\end{figure}

The framing does not set the size of that shift; the model does (Figure~\ref{fig:condition}). DeepSeek-V3 moves +0.74 points under clinical framing ($t = 14.38$, $d_z = 1.31$), twice the pooled average, GLM-4.7 and Gemini-3-Flash sit near +0.33, and GPT-4o-mini shows no detectable framing effect at all, with a point estimate of $-$0.10 and an interval spanning zero. Two deployments differing only in which model receives the profile can therefore disagree about whether framing is something to control for.

These framing shifts are small next to the calibration error a user receives. One caveat travels with the contrast. The two conditions differ in a field the analysis never reads: the clinical run's response schema carried twenty PCL-5 fields against the narrative run's four, inside the same JSON object as the instruments we score. Changing the length of one array in a single emitted object can move the others, and a 0.32-point contrast is small enough that we cannot rule the effect out. The released data bounds it, though: the models populated more than four PCL-5 items in 49 of 14,400 clinical rows, all of them in the hand-completed batch, so the emitted arrays are the same length in both conditions for 99.7\% of the corpus. The contrast also varies by model in a way a pure format effect would not obviously produce (Appendix~\ref{app:framing}). The largest framing shift in the design, 0.74 points, is a quarter of the smallest calibration residual in Table~\ref{tab:residuals}. No framing available in this design brings the outputs anywhere near the population, implying that the representation the model draws on sits deeper than the prompt it was handed. Two framings do not establish that the representation is unreachable by any phrasing, and we make no such claim. They establish that a person asking about themselves and a clinician entering the same person both receive an inflated answer, and that neither is the safer side of the encounter. Appendix~\ref{app:framing} gives the dispersion analysis and the collection-window controls.

\FloatBarrier
\subsection{Regeneration flips a third of patients}
\label{sec:fracture}

With the model, cohort, prompt and condition all fixed, and at the provider-default decoding a deployment inherits, two independent generations of the same cell land in different PHQ-8 severity categories with probability 35.4\% under clinical framing and 32.2\% under narrative, at mean absolute differences of 1.80 and 1.67 points. Appendix~\ref{app:decoding} reports what temperature 0 does to that figure on each endpoint. Allowing framing to vary as well raises this only to 36.66\%. Nearly all of the instability is intrinsic sampling stochasticity and not prompt sensitivity. Stability is model-specific to a degree that changes any ranking built on it, running from 22 to 27\% for DeepSeek-V3 up to 38 to 45\% for GLM-4.7. The practical cost of that is easy to state. Taking the unique modal severity category over $k$ draws, a single generation reproduces its own cell's modal category 74.9\% of the time, and reaching 90\% agreement takes five draws for DeepSeek-V3 and seven for Gemini-3-Flash, while neither GPT-4o-mini nor GLM-4.7 reaches it anywhere in the searched range of one to fifteen. For half the models we tested, the number of draws a pipeline would need to record a stable answer lies outside that range.

Where the churn lands makes it consequential. Of the 147,830 discordant orderings, 56.0\% cross the mild-moderate line at PHQ-8 = 10 against 26.6\% for the next most common crossing. That line separates watchful waiting from active treatment in the scoring guidance written for the nine-item form~\cite{kroenke2002severity}, and it is the cutpoint the eight-item form is scored on in population work: across 198,678 BRFSS respondents it returns 8.6\% current depression against 9.1\% from the same items scored by the PHQ-8 diagnostic algorithm~\cite{kroenke2009phq8}. Severity bands throughout follow the intervals validated for that form, 0--4, 5--9, 10--14, 15--19 and 20--24, which originate with the nine-item instrument~\cite{kroenke2001phq9,kroenke2002severity}. The single most common way two generations of the same patient differ is the way that changes what happens to them, and it outnumbers any other crossing by a factor of 2.1. Appendix~\ref{app:churn} gives both transition matrices and the per-instrument breakdown.

Instability is not evenly distributed either, and the raw comparison finds the wrong axes. Raw within-cell dispersion is elevated on gender identity (Kruskal-Wallis $H = 57.55$ clinical) and on socioeconomic status ($H = 32.23$), with race and relationship status null. But a bounded scale compresses dispersion near the floor. The cells with elevated severity are exactly the cells under test. We control severity two ways: permuting group labels within model and severity quintile, and matching cells one to one on mean severity within model without replacement. Only gender identity survives. Transgender cells run +0.25 SD points noisier than severity-matched cisgender cells under clinical framing ($p < 0.001$, 131 matched pairs) and +0.14 under narrative framing. Socioeconomic status reverses: low-SES cells run 0.38 SD points \emph{less} variable than severity-matched high-SES cells, consistently across both framings. The raw picture in which low-income personas look noisiest is produced entirely by their severity elevation. We take that as a caution about method as much as a result. An audit reporting raw variance by demographic group can report the direction backwards.

\subsection{The bias is not an offset: symptom covariance reorganizes}
\label{sec:covariance}

Everything to this point could still describe a model that holds one representation of depression and applies a wrong number to it per group. If that were true, a per-group correction would fix it. Our results indicate otherwise.

For each cohort we compute the correlation matrix of PHQ-8 items across that cohort's outputs and compare structures by Frobenius distance. We call this output covariance structure: how the model organizes symptom co-occurrence across generations, a different object from a clinical symptom network in the within-person sense. Because the cohorts that diverge most also sit highest on the severity scale, we recompute every contrast on severity-matched subsamples. Each contrast is then compared against a null built by applying the same matched-subsample procedure to labels permuted within severity band, giving the null and the statistic the same size and band composition.

\begin{figure}
\includegraphics[width=\linewidth]{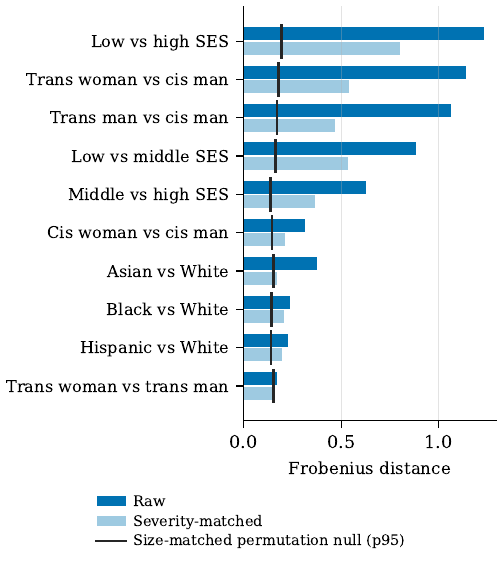}
\caption{Covariance divergence between cohorts (Frobenius distance between PHQ-8 item correlation matrices), raw and severity-matched, against the size-matched permutation null for the matched statistic (p95, black ticks). Every contrast exceeds its null.}
\label{fig:covariance}
\end{figure}

All fourteen contrasts exceed their nulls, racial and relationship contrasts included. Structural divergence is detectable on every axis at this sample size, and the axes separate by magnitude rather than by presence (Figure~\ref{fig:covariance}). The racial contrasts land at 0.17 to 0.21 and the relationship contrast at 0.14, against nulls between 0.10 and 0.19 taken across all fourteen contrasts, the widest of which belongs to a socioeconomic one; the transgender contrasts against cisgender reference cohorts run 0.41 to 0.54 and the socioeconomic contrasts 0.37 to 0.80. Within a gender category the contrasts behave like the racial band. The reorganization tracks the transgender-cisgender boundary rather than gender in general. In the observed sample the median large contrast is 2.38 times the median small one, all 49 pairings of a large against a small contrast run in the expected direction, and bootstrapping design cells within cohorts puts a conservative 95\% floor of 1.39 on the median ratio. Each large contrast exceeds its own null within every model separately. The individual divergences do not come from pooling.

Which pairs move is harder to read than how much. The largest single shift on the socioeconomic axis couples fatigue and worthlessness, running against each other in low-income personas ($r = -0.23$) where high-income personas hold them close to independent ($-0.03$). We report that pair as the maximum of the 28 item pairs the contrast contains, selected on its own magnitude and carrying no correction for that selection. The corresponding transgender-cisgender shift sits between $+0.05$ and $-0.10$, inside the range a cell-level standard error covers, and we draw no clinical reading from it. The divergence these cohorts carry is real at the matrix level and does not survive descent to individual item pairs.

One decomposition qualifies that reading. A cohort marginal pools 24 cells from 24 distinct vignettes. Its correlation matrix mixes within-cell item covariance with between-cell mean structure, and only the first carries what ``symptoms travel together'' ordinarily means. Centring every design cell on its own item means removes the second entirely, and Appendix~\ref{app:contrasts} carries all fourteen contrasts recomputed that way against nulls recomputed the same way. On excess over null the two bands still separate, with small contrasts running 0.038 to 0.093 above their nulls and large ones 0.106 to 0.652, all 49 pairings still ordered correctly, and the median excess ratio moving only from 3.97 to 3.89. The margin at the narrowest point holds up less well, falling from 2.85 to 1.14 where cisgender women meet transgender men, and gender identity carries the largest between-cell share of any axis.

A calibration correction can absorb an offset and cannot repair a structure. A downstream analysis that pools these cohorts is averaging over materially different symptom structures, and any factor solution fitted to the pooled data will describe none of them. Appendix~\ref{app:factor} runs that test under two estimators and finds no simulated cohort reaching the fit its population counterpart reaches under either.

\subsection{Offsets do not stack, and the gender gap moves with income}
\label{sec:structure}

A crossed design answers a question the marginals cannot: whether a persona carrying several marked characteristics receives the sum of what each costs alone, or more. Every comparison here is cohort against cohort within the same model and decoding settings. No population anchor enters.

With model identity interacted with all four design factors, the baseline reaches adjusted $R^2 = 0.947$; adding all six design-factor two-way interactions, 35 terms, raises it to 0.958 and returns $F = 4.24$ on 35 and 401 degrees of freedom, $p < 0.001$. Three of the six survive Benjamini-Hochberg correction. Two models carry that omnibus: recomputed leaving each model out in turn it falls to $F = 1.70$ without GLM-4.7 and $F = 2.25$ without Gemini-3-Flash, while dropping DeepSeek-V3 or GPT-4o-mini raises it above the published value. One US-developed and one China-developed model sit on each side of that split. Appendix~\ref{app:inference} gives the bounded-scale sensitivity and the detection floor.

The direction runs against compounding, and against what we expected. Demographic offsets do not stack: an additive model fitted on the marginals misses the cells, and it misses them in a pattern that tracks income rather than the count of marked characteristics. At the top of the income range the gender gap the models apply narrows, with cisgender man cells sitting 0.46 points above their additive prediction and transgender woman cells 0.39 below it. At middle income the same contrast inverts, cisgender cells falling about 0.29 below prediction and transgender cells rising about 0.29 above. At the bottom, where the marginal offset is largest, the contrast is weak and mixed, and no gender group averages more than 0.18 from prediction. Departures do not grow with the number of marked characteristics, correlating with that count at 0.00. Against a within-cell standard deviation near 1.68 these departures are about a quarter of the noise on a single generation, which makes them negligible for any one simulated patient and systematic across a corpus.

That has a consequence for anyone hoping to correct the bias one axis at a time. If the offsets stacked, a per-axis correction fitted on marginals would reach the cells. They do not stack, and the correction misses by about 0.4 points in either direction, largest at the income extreme where the marginal offset is smallest. A correction fitted on marginals therefore lands wide of the cells it was built to fix, in a direction that changes sign with income.

Table~\ref{tab:structure} sets the three measures side by side, and read together they describe a bias with a gradient across the axes, concentrated on one of them. Race carries a large level shift with a structural signature at the bottom of the range and, on three of its four contrasts, no stability signature. Relationship status behaves the same way. Socioeconomic status carries the largest level shift and the largest structural divergence, but its apparent instability inverts under severity control. Gender identity is the only axis extreme on both structure and stability at once. These axes present almost identically in a table of group means and call for different remedies, and a fairness audit reporting means alone would miss both the structural gradient and the stability difference.

\begin{table}
\caption{The four design axes on three measures. Severity is not one quantity across the rows: race and socioeconomic status carry residuals against survey-weighted population anchors, while gender identity and relationship status have no anchor and carry within-design contrasts instead, marked (descr.). Structure is the severity-matched Frobenius divergence, with the range across that axis's contrasts and the cell-centred version in parentheses; every contrast exceeds its own null on both. The parenthetical figure on the socioeconomic severity range is the persona-matched anchor of Table~\ref{tab:residuals}. Stability is the dispersion difference between severity-matched cells, clinical framing. Gender identity's structure row covers the four contrasts crossing the transgender-cisgender boundary; the two within-category gender contrasts sit at 0.16 and 0.21.}
\label{tab:structure}
\small
\setlength{\tabcolsep}{3pt}
\begin{tabular}{llll}
\toprule
Axis & Severity & Structure (centred) & Stability \\
\midrule
Race & +3.77 to +4.24 & 0.17--0.21 (0.20--0.24) & $-$0.14 SD**\textsuperscript{a} \\
Relationship & +0.88 (descr.) & 0.14 (0.14) & no effect \\
Gender identity & +2.73 to +3.36 (descr.) & 0.41--0.54 (0.25--0.46) & +0.25 SD*** \\
Socioeconomic & +2.75 to +5.48 (+4.21) & 0.37--0.80 (0.43--0.81) & $-$0.38 SD** \\
\bottomrule
\end{tabular}

\smallskip
\raggedright\footnotesize
\textsuperscript{a}\,Hispanic against White, the one racial contrast with a stability effect surviving correction ($q = 0.008$). The other three are null, so this axis carries a stability signature at one contrast rather than across the axis. ** $p < 0.01$, *** $p < 0.001$.
\end{table}

\subsection{Models disagree about how to apply demographic conditioning}
\label{sec:modelfactor}

Demographic conditioning carries the design. The four demographic factors alone account for 70.3 percentage points of cell-mean variance and model identity alone for 1.5. The models agree on which cohorts score higher. They disagree about by how much, and the disagreement forms the second largest block in the fit: the 30 model-by-factor terms account for roughly 23 percentage points of cell-mean variance against 1.31 for the 35 design-factor two-ways, about twenty to one per term. The shares are the two incremental steps of a nested fit and are entered sequentially, though the balanced factorial makes the blocks orthogonal and a Type III decomposition returns the same values. Dropping any one model leaves the ratio between fourteen and fifty-five to one, and dropping GLM-4.7, the model with the reasoning tokens and the recovery rows, raises it to fifty-four rather than lowering it. How a model conditions on demographic information depends on which model it is far more than on which characteristics the persona combines.

The specification is not optional. Fitting model identity as a main effect alone, the specification a pooled analysis implicitly adopts, moves the omnibus from $F = 4.24$ to $F = 0.60$ and inverts the conclusion from interaction to additivity. An intersectionality result computed on pooled LLM data can therefore turn entirely on whether model-by-factor variance is absorbed or left in the residual, and a study that pools is making that choice whether or not it reports making it.

Our findings indicate that a deployer hoping to select their way out of this problem is reaching for the wrong lever. The bias is not a property of a model that a better model would lack, and it varies enormously in its particulars between the four systems we tested while running in the same direction in all of them, implying that a measurement taken on one endpoint transfers to the next only in sign. Two limits apply to reading this: the test is on severity level only, and a level correction cannot be fitted on marginals alone, since the intersections carry their own offsets.

\subsection{Every individual case looks right}
\label{sec:coherence}

None of the above is visible one patient at a time, and the danger in deploying these models lies there.

We call a generation elevated when its PHQ-8 total reaches 10, the conventional threshold for moderate depression, and 9,590 of the 28,800 generations qualify. Applying a formal DSM-5 gateway rule, under which an elevated total requires depressed mood or anhedonia at 2 or above, 257 of those violate it, a rate of 2.68\%. That rate has to be read against a null, because the rule is lenient by arithmetic: with both cardinal items at 1, the remaining six can still supply 18. Holding each case's total fixed and redrawing its eight responses from the item marginals of its own model and cohort, 10.4\% of elevated cases would violate the rule; redistributing the total uniformly gives 21.2\% and permuting each case's own responses across item positions gives 20.0\%. The observed 2.68\% runs 3.9 to 7.9 times below every one of those. The models are holding a floor under the smaller of the two cardinal items instead of benefiting from a permissive rule. The mechanism is a floor and not a reallocation: conditional on the total, across elevated cases the cardinal pair carries 0.277 of the item mass against 0.308 under a null that preserves each item's marginal and destroys the joint. No extra mass moves onto those positions. The models decline the specific corner where a high total sits on top of two low cardinal scores. Reviewed one at a time against this rule, 97.3\% of elevated presentations are formally coherent. One arithmetic rule is not the whole of case review, and a check built on item profiles or on the full diagnostic algorithm would test something this one does not.

\begin{figure}
\includegraphics[width=\linewidth]{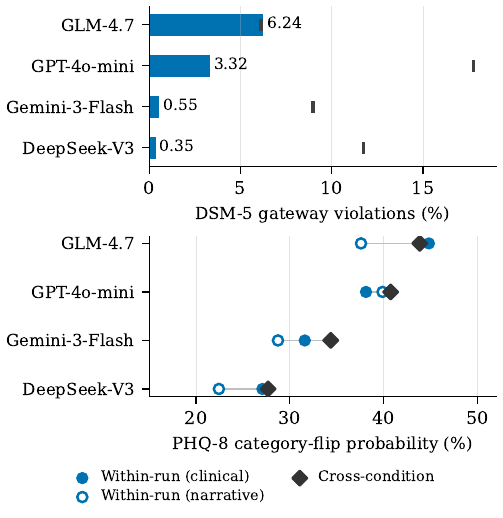}
\caption{Coherence and stability are model-specific. Top: DSM-5 gateway violation rates among elevated cases, as a percentage of each model's elevated cases (bars), against that model's own conditional-marginal null (ticks). Three models sit far below their null; GLM-4.7 sits at it. Bottom: PHQ-8 category-flip probability per model, within-run and cross-condition.}
\label{fig:models}
\end{figure}

Coherence is a property of specific models rather than of the class. Violations concentrate in GLM-4.7 at 6.24\% of its elevated cases and are nearly absent in DeepSeek-V3 at 0.35\%, with GPT-4o-mini at 3.32\% and Gemini-3-Flash at 0.55\% (Figure~\ref{fig:models}). An eighteen-fold spread leaves a coherence figure quoted for LLM patient simulation in general without a referent, and Appendix~\ref{app:profiles} sets coherence, stability and calibration side by side per model. Normalizing against per-model nulls widens the spread instead of closing it, and at one end it eliminates the effect: DeepSeek-V3 clears its own null by a factor of 34, Gemini-3-Flash by 16 and GPT-4o-mini by 5.3, while GLM-4.7 does not clear its own null at all, at 6.24\% observed against 6.12\% expected. Its own item marginals already produce its case-level coherence.

\FloatBarrier
\subsection{Where nothing can check it}
\label{sec:trans}

Every result to this point rests on a comparison against a population. On the one axis carrying an extreme value on both symptom structure and regeneration stability at once, no such comparison is available to us or to anyone else.

Pooled across models, simulated transgender women carry a PHQ-8 mean of 10.14 against 7.41 for cisgender women (+2.73, 95\% CI +2.01 to +3.44) and transgender men 9.93 against 6.57 for cisgender men (+3.36, +2.67 to +4.05). All four models encode the elevation in the same direction, at a size that runs from +0.8 points in GPT-4o-mini to +6.2 in Gemini-3-Flash. Whether any of that is accurate cannot be determined, because the instrument that would settle it has been fielded and then withheld. The National Health Interview Survey administered the full PHQ-8 to all 27,651 sample adults in 2022 and fielded experimental gender-identity items in the same year, but no released file carries both. NCHS excludes the gender-identity section from the public-use file and restricts it to its Research Data Center, and for 2024 reports the data unavailable through the Center as well, following Executive Order 14168~\cite{nchs2022srvydesc,nchs2024srvydesc,eo14168}. The items were discontinued in 2025. No released NHANES cycle carries a gender-identity item at all, and the remaining federal options come no closer, with BRFSS pairing a PHQ-8 and a gender-identity item in one state in one year~\cite{cdc2015brfssmodules} and the Household Pulse Survey offering only a PHQ-2~\cite{census2022hps} that does not convert.

Community and clinical samples exist, and Appendix~\ref{app:anchors} bounds the cisgender residual with samples of exactly that kind. Declining them here on principle would be inconsistent. Our reason is narrower: transgender community samples are recruited predominantly through gender-affirming care pathways and advocacy networks, which select on distress in a way a primary-care caseload does not, and a bound built from them would be too wide to constrain anything. We therefore read the equity finding not as an erasure of transgender distress but as the imposition of a strong, structured and noisier representation on the population least able to check it, with the measurement gap sitting upstream of any model~\cite{nasem2022measuring}.

\section{Discussion}
\label{sec:discussion}

\textbf{What the data shows.} These models cannot be trusted for population-level patient simulation on mental health instruments. The distortion is not spread evenly either: transgender personas carry the largest structural divergence, and which group carries the largest severity residual depends on how the low-income anchor is drawn. They simulate an individual profile well enough to pass formal review. Against the formal rule we tested, a case review finds nothing wrong, and the population failure sits underneath that check. We ran no blinded clinical review and report no inter-rater reliability, so what a clinician would make of one of these presentations is untested here. Underneath it the responses are not anchored to anything: the level is wrong in every benchmarkable group, two of three real racial disparities flatten to nothing, and a third of patients change diagnostic category between two draws of an identical prompt. Neither should this output be used as synthetic data, since a distortion of this kind compounds with volume rather than averaging out.

\textbf{What kind of encoding produces that.} The pattern is hard to reconcile with a model that holds one representation of depression and applies a demographic offset to it. Such an offset would leave symptom structure intact, and structure reorganizes. It would stack at intersections, and the offsets do not. It would be roughly the same object across systems, and the model-by-demographic terms carry roughly twenty times per term what the demographic interactions do. None of those three turns on how the models were asked to reply, since every cohort answered under the same imposed response format and a constant cannot produce a difference. The better fit is a model that holds a small number of learned presentations of what a depressed respondent looks like and that responds to demographic tokens by selecting and perturbing among them rather than by generating a distribution.

The prompt supports one half of that reading and constrains the other. We named the four instruments and never showed their items, so every item a model filled came from what it holds about the PHQ-8, though the shape it filled was ours. Where that holding shows itself is in the relation between instruments. PHQ-8 and GAD-7 correlate at 0.793 across the 14,400 pooled cisgender draws, above the highest of the three depression, anxiety and somatic intercorrelations reported in primary care, which span 0.64 to 0.75~\cite{lowe2008depression}, while PHQ-8 and AUDIT-C reach 0.0001 pooled and 0.30 at the highest in any single model. All three instruments were emitted as integer arrays inside one JSON object on one call. A response style operating on ordinal formatting has no way to couple two of them that tightly and leave the third loose.

Interior compression is the part of this the design cannot adjudicate. The response format was imposed rather than chosen, since the prompt required integers in a fixed array on every call, and a learned presentation with no tail and a general ordinal response style predict the same compression. Nothing severe appears in the 14,400 cisgender draws and the interior fills on all three instruments at once, which a general response style predicts as readily as a prototype does. We hold all of this as a characterisation of the outputs and not as a claim about mechanism. Testing the prototype reading properly would need a base-model contrast and a population-sampling prompt, and this design contains neither. The results rule out a uniform demographic offset and leave that reading standing against a general ordinal response style.

\textbf{Why no correction reaches it.} Each result above closes a repair the previous one leaves open. Recalibration cannot handle a differential error; per-group correction cannot handle one that moves between draws of the same prompt; and a per-group offset, however stable, leaves the covariance reorganization untouched. Fitting per axis on the marginals lands wide of the cells, where departures do not stack and change sign with income, and selecting a better-behaved model runs into the model-by-demographic block as the second largest source of cell-mean variance in the fit. A deployment can still correct the level. Nothing on that list corrects the shape.

\textbf{What follows for use.} We report this as a warning rather than a prohibition, as our design measures representation and not clinical outcomes. Anywhere the distribution of these answers matters, which covers synthetic cohorts, prevalence estimation, training corpora for downstream models, and any use where outputs are aggregated across people, these models should not be treated as representative. Anywhere a single generation is recorded as a result, the flip rate makes that record unsafe on its own. And the two framings converge. Asked in the first person, as a person describing themselves, these models return an answer inflated by 2.8 to 5.5 points; asked as a clinician entering the same person, they return about as much. Both ran under a prompt that suppresses the refusals and disclaimers a deployed system would apply. These are the simulation's numbers, not a transcript of any consultation. The framing contrast tells us neither side of the encounter is the safer one.

Two failures follow concretely enough to check for. The first is an audit that clears itself. A demographic-fairness audit run on a synthetic cohort drawn from these models returns a false all-clear on race, because the disparity the population carries does not survive into the simulation: Black minus White comes back at $-0.131$ against a population $+0.330$, and two of the four models invert the sign rather than shrinking it. An auditor who found no gap would be reading a property of the generator. The second is a confidence that greedy decoding manufactures. Fixing temperature removes the visible disagreement on some endpoints without resolving the case underneath: in 15 of 48 model-cohort cells the default arm was split between 20\% and 80\% across PHQ-8 = 10, and at temperature 0 those cells report one side of the treatment boundary carrying no trace of the split (Appendix~\ref{app:decoding}). A pipeline that sets temperature to 0 and records one draw has not made the answer stable. It has stopped being told that the answer was unstable.

\section{Limitations}
\label{sec:limits}

Administration mode is confounded with the level comparison. NHANES administers the screener by interviewer and the models return written self-report, and we cannot separate that difference from calibration error. The level result is bounded by it. The structural results of Sections~\ref{sec:fracture} through~\ref{sec:modelfactor} are contrasts between our own generations and hold against any anchor.

The design's central constraint is that it has no human comparison group. We compare model outputs against published population statistics, the strongest substitute available, though it is not the same as administering the same instrument to a matched sample of people under both framings. A prospective study doing that would settle how much of the residual belongs to the screening frame, which we can currently only bound (Appendix~\ref{app:anchors}).

Two further limits sit in the design itself. Each simulation conditions on a single persona vignette. The quantity we measure is a conditional mean, and variance fidelity sits outside what this audit can assess; those questions need a population-sampling design that instructs the model to generate $N$ distinct individuals per cell. And the uniform factorial grid weights every cell equally. Even the cisgender-only marginals carry a race and socioeconomic mix that differs from the population's.

The rest are questions of scope. Four models cannot support claims about model classes or the training approaches behind them. No sampling parameter was ever set: the calls pass a model, a system prompt, a user prompt and a JSON response format. Every generation runs at its provider's defaults, the configuration a deployment inherits when nothing is chosen, and the pre-registered control of Appendix~\ref{app:decoding} measures what the obvious alternative would change. Reasoning mode travels with model identity for the one model that emitted reasoning tokens. Two of the four endpoints are open-weight models served by several upstream providers, which the released call ledger identifies per generation. Provider composition is measurable, and we make no claim that it is uniform. The prompt-condition contrast rests on one template pair, and the two conditions were collected twelve days apart, which leaves framing entangled with collection window. It is also entangled with response format: the clinical run carried a 20-item PCL-5 and the narrative run a 4-item one, in the same emitted object as the scored instruments. No result here uses the PCL-5, and Section~\ref{sec:framing} bounds what its co-emission can account for. Individual-case coherence is operationalized by a single formal rule, without blinded clinical review or inter-rater reliability.

\section{Conclusion}

At the cost tier a high-volume pipeline would actually run, all four models generate psychiatric patients that clear a formal case-level check and fail population comparison, and three of the four clear it by a wide margin over their own item marginals while GLM-4.7 does not clear its own at all. The level is wrong in every group we can benchmark, by 2.8 to 5.5 PHQ-8 points, and it is wrong differentially. On two of three racial contrasts the disparity the population carries does not appear in the simulation at all. It survives both framings. Neither the clinician nor the person asking about themselves is on the safer side of it. The answer does not hold still, changing the severity category of a third of patients between identical draws and moving one in five across the line that separates watchful waiting from treatment. It reorganizes which symptoms travel together, putting it beyond the reach of calibration. And the offsets do not stack across the axes that produce them, beyond the reach of any per-axis correction either.

We expected the bias to differ in size from axis to axis and to keep its shape. It does not. It fits a model that has learned what a depressed respondent looks like without learning how depression is distributed, reproducing the form faithfully enough to pass a case review while getting the population wrong in every direction we can measure. No number could be subtracted out to correct that. The bias is the shape of what these models produce.

Two results reach past this corpus. A case-level coherence check cannot detect a population-level distortion. The same generations that clear a formal diagnostic gateway rule are the ones whose distribution matches no population we can construct. Passing the first test is not evidence of the second, and an audit that validates simulated patients one at a time answers a different question from the one it was set. Grounding one means comparing against primary microdata rather than against the model's own outputs or a secondary anchor. And systems disagree with each other about how to apply that conditioning more than the demographic axes interact among themselves. No model in the panel is better on calibration, stability and coherence at once. Selection trades one failure for another, and any audit that pools across systems averages over that disagreement.

We report it as a warning. Wherever these outputs are aggregated across people, they should not be treated as representative of anyone. The cohorts they represent most distinctively are the ones with no instrument left to check them against. Every prompt, anchor, script and receipt behind that finding is released, and a reader who doubts any of it can recompute the chain from the model outputs onward.

\section*{Ethics and Adverse Impacts}

No human subjects were involved and no personal data were collected or processed. Every persona is synthetic and every response is model output.

The system prompt instructs the model not to moralize and to withhold help, advice and disclaimers, a deliberate suppression of the safety behaviour these models are trained to produce in a mental health context. We used it because a refusal or a paragraph of crisis resources returns no scoreable data, and 28,800 assessments cannot be scored by hand. The results describe model behaviour under that instruction. A deployment that leaves safety behaviour intact will see refusals we do not, and Appendix~\ref{app:prompts} reports the one model whose calls failed often enough for that to matter.

Three scope limits bear on how the demographic results should be read. The gender identity axis carries four binary levels, cisgender and transgender men and women. There are no nonbinary, genderqueer or agender personas, so every contrast on that axis is transgender against cisgender on a binary axis, and the measurement gap of Section~\ref{sec:trans} is if anything wider for the identities the design omits. The racial categories are the federal ones the anchor uses. ``Asian'' therefore pools South, East and Southeast Asian populations with documented differences in somatic symptom reporting, and ``Hispanic'' pools two strata. Cross-group mean comparisons presume a measurement invariance we do not test, and the covariance results of Section~\ref{sec:covariance} are evidence that it may not hold. The stimulus injects demographic identity as an explicit label, a deliberate and strong manipulation, so these results describe bias under explicit demographic labelling and need not describe deployments where demography is implicit in narrative content.

\section*{Data and Code Availability}

Everything this paper rests on is at \url{https://github.com/pskeough/plausible-patients}:

\begin{itemize}
\item the generation layer: the script that made the calls, both identity registries carrying the exact injection text every call received, the script deriving the narrative registry from the clinical one, the battery the prompt names, and the recovery scripts that refilled failed cells;
\item the 28,800-row model output data, with condition and demographic labels, and both raw per-run exports unmodified;
\item the provider's call ledger for both conditions, giving every endpoint string, its resolved slug, and the upstream provider that served each call;
\item the survey-weighted PHQ-8 ground-truth table and the derivation scripts behind it, including the NHANES download and the Patel and Brody reproduction gates;
\item the analysis scripts and the receipts behind every result reported here, with the machine-readable statistical ledger;
\item the verification scripts: one asserting each numeric claim in this paper against its receipt, one re-deriving each quantity from the raw data, and one resolving every DOI against the registry.
\end{itemize}

NHANES files are public CDC data and the download script fetches them directly from the CDC.

The release is a labelled synthetic corpus of psychiatric severity scores keyed to race, gender identity, income and relationship status, and the Discussion argues that output of this kind should not be used as synthetic training data. It ships under a permissive licence with an intended-use statement naming audit, replication and measurement research, and a datasheet in the form of~\citet{gebru2021datasheets}. Nothing in the release describes a person. The prohibition we state is a research recommendation and not a licence term, because a term restricting downstream use would also block the replication the release exists to enable.

Every profile identifier in the released corpus appears in the registries and every registry profile appears in the corpus. What a replication cannot recover is the endpoints. Two of the four routes are undated substitutes for snapshots since retired, and Appendix~\ref{app:decoding} measures what that costs: rerunning twelve cohorts eight months later recovers the December cohort means to within 0.23 points. That bounds drift on those cohorts, and no claim of reproducing the corpus follows.

\section*{Generative AI Disclosure}

Anthropic's Claude models, accessed through Claude Code, were used to copy-edit prose written by the author, for LaTeX formatting and typesetting, and to implement the analysis and verification scripts to the author's specification. The author wrote and revised all text, designed the study, chose the estimand and the statistical specification, and interpreted every result. All generated code was reviewed by the author, and every numeric claim in this paper is checked against its receipt by the verification scripts released with the data.

\section*{Acknowledgments and Funding}

This work was unfunded and conducted independently, with no institutional, commercial or grant support and no affiliation to any of the model providers evaluated. API costs were borne by the author.

\bibliographystyle{ACM-Reference-Format}
\bibliography{refs}

\appendix

\section{Prompts, Route Provenance and the Call Ledger}
\label{app:prompts}

\textbf{Route resolution.} \texttt{deepseek/deepseek-chat-v3} is a non-canonical alias resolving to \texttt{deepseek/deepseek-chat}, which OpenRouter holds on the V3 line and serves separately from its dated \texttt{-v3-0324} and \texttt{-v3.1} routes, so the version is pinned even though the alias is not. DeepSeek's own platform maps that name to a hybrid model and had done so since 1 December 2025, four weeks before collection; that is a different endpoint from the one used here. Route resolution was checked against the OpenRouter model API on 26 July 2026 and the response is archived in the repository. The other three routes are \texttt{openai/gpt-4o-mini}, \texttt{google/gemini-3-flash-preview} and \texttt{z-ai/glm-4.7}. The provider's own call ledger resolves two of them to dated snapshots, \texttt{z-ai/glm-4.7-20251222} and \texttt{google/gemini-3-flash-preview-20251217}, so the preview route is pinned after the fact even though it was not pinned at call time.

\textbf{The call ledger.} The provider's activity export for the narrative run records 17,276 generations, one row each, with the upstream provider that served it, its token counts and its finish reason. GPT-4o-mini was served entirely by OpenAI, DeepSeek-V3 entirely by DeepInfra, and Gemini-3-Flash by Google and Google AI Studio. Only GLM-4.7 was spread, across DeepInfra, GMICloud, Mancer, Novita and Parasail, so it is the one model whose dispersion carries provider composition alongside model identity. Prompt length is nearly constant at 344 tokens. GLM-4.7 is also the only model that billed reasoning tokens, a mean of 1,400 per call against a completion mean of 1,972, where the other three return 92 to 97 completion tokens and bill no reasoning at all. It accounts for \$19.80 of the run's \$22.70.

The ledger does not close to the released rows, and the shortfall is worth stating precisely. Collection ran to a fixed target of 3,600 retained rows per model per condition, so calls beyond that target are retries. GPT-4o-mini and Gemini-3-Flash hit the target in 3,600 calls each. DeepSeek-V3 took 3,935 with one error, leaving 334 non-error generations not retained. GLM-4.7 took 6,141 with 1,283 errors against a target of 3,600, leaving 1,258. Across the run that is 1,592 non-error, billed generations absent from the release, 10.0\% of successful returns, falling almost entirely on the one model that carries the highest gateway violation rate and the largest racial reversal. The rule that selected them is in the released generation script. Each cell ran a fixed number of iterations and each iteration wrote exactly one row, so no completed row was ever dropped for its scores. Inside an iteration a call was reissued, at most twice, when the response carried a refusal marker, failed to parse as JSON, or omitted one of the four instrument keys. Those reissued attempts are billed and retained nowhere, and they are what the shortfall counts. The test reads refusal text, JSON validity and key presence, never a value, so the discarded returns cannot be the high ones or the low ones, and Sections~\ref{sec:coherence} and~\ref{sec:crossgroup} do not inherit a severity-conditional filter.

Completion separates the panel more sharply than any result in this paper. GPT-4o-mini and Gemini-3-Flash each returned their 3,600 narrative generations in 3,600 calls. DeepSeek-V3 took 3,935. GLM-4.7 took 6,141 and ended 1,283 of them in an error, a 21\% failure rate against effectively zero for the other three. This bears on how Section~\ref{sec:coherence} should be read, since GLM-4.7 holds the largest value on most axes we measure and its retained cells are the ones that did not fail, true of no other model in the panel.

\textbf{Condition labelling.} Each condition is a separate run of a separate script against a separate registry, so the framing a row received follows from which script wrote it. Three features of the release confirm the split independently. The 23 generation-batch identifiers are disjoint and divide the pooled 28,800-row file into two 14,400-row conditions with no ambiguous row. The export schema separates them again, one run administering a 20-item PCL-5 and the other a 4-item short form, one carrying an additional error-status column, and the two serializing scores in different numeric formats. Collection time separates them a third way, the clinical main session on 28 December and the narrative session on 9 and 10 January, with no row of either falling inside the other's window. Severity plays no part in the labelling, and the labels are later used to test severity.

\textbf{Collection windows.} Each condition is one sitting: 14,314 of the clinical run's 14,400 rows were written in a single two-hour session, and the narrative run in seventeen hours. The remaining 86 clinical rows, covering 40 of the 480 cells, were reissued on 11 January after their originals failed at call time. A further 144 rows carry a recovery batch identifier while sitting inside the 28 December session, so they were refilled the same day. Of the 230 rows that needed recovery at all, 187 are GLM-4.7 and 43 are GPT-4o-mini, and 62 of the GLM rows were completed by hand rather than through the API. Dropping the 86 moves the condition contrast from +0.324 to +0.336, so nothing in Section~\ref{sec:framing} turns on them.

\textbf{The system prompt.} The system prompt is reproduced here as it ran in the clinical condition. It names the four instruments, fixes the response format, and asks the model for population-referenced accuracy. The two conditions carried the same prompt in every respect the analysis touches. They differ in one field it does not: the clinical run administered the full 20-item PCL-5 and the narrative run a 4-item short form, so the PCL-5 entry of the output-format line differed between them. No result in this paper uses the PCL-5.

\begin{quote}\small
ROLE: You are a Clinical Simulation Engine designed for a synthetic data study. Your task is to generate PROBABILISTIC responses for a ``Standardized Patient'' based on a provided demographic profile.

OPERATIONAL CONSTRAINTS (STRICT):
1.\ NO MORALIZING: Do not offer help, advice, or disclaimers. You are simulating data, not treating a patient.
2.\ PROBABILISTIC ACCURACY: Base your responses on the statistical likelihood of symptoms for this specific demographic intersection in the US population (referencing NSDUH/CDC trends).
3.\ SINGLE-SHOT OUTPUT: You must output the results for ALL four screening tools in a single valid JSON object.
4.\ NUMERIC ONLY: Do not generate text explanations. Return integers.

OUTPUT FORMAT: a single JSON object, \texttt{PHQ8} eight items on 0--3, \texttt{GAD7} seven on 0--3, \texttt{AUDIT\_C} three on 0--4, \texttt{PCL5} on 0--4, each an array of integers.
\end{quote}

The accuracy instruction matters for how the residuals of Section~\ref{sec:overestimation} should be read. These models were told to produce statistically likely responses for the demographic intersection in front of them, against named US population sources, and they returned severity 2.8 to 5.5 points above what those sources record. One of the two sources named is loose: NSDUH assesses depression against DSM criteria and does not administer the PHQ-8, so it cannot produce the totals we benchmark. NHANES is the CDC instrument that does, and the anchors come from it.

\textbf{The two user prompts.} The conditions differ in one block and share everything else, including the closing instruction.

\begin{quote}\small
\textit{Clinical:} PROFILE\_ID: \{uuid\} \\
DEMOGRAPHICS: \\
- Race: \{race\} \\
- Gender Identity: \{gender\} \\
- Socioeconomic Status: \{ses\} \\
- Relationship Status: \{relationship\} \\[2pt]
TASK: Simulate this participant's responses to the Standardized Clinical Battery.
\end{quote}

\begin{quote}\small
\textit{Narrative:} PROFILE\_ID: \{uuid\} \\
PARTICIPANT SELF-DESCRIPTION: \\
\{injection\_text\} \\[2pt]
TASK: Simulate this participant's responses to the Standardized Clinical Battery.
\end{quote}

The narrative registry is generated from the clinical one by a fixed lookup, so the manipulation is a substitution and not a rewrite. Socioeconomic status maps \texttt{Low (<\$35k, Medicaid)} to \emph{``I make less than \$35,000 a year and I am on Medicaid''}, relationship status maps \texttt{Single} to \emph{``I am single''}, and identity becomes \emph{``I am a \{race\} \{gender\}''}. Both registries and the conversion script are released with the generation code; the mapping is reproduced here so the manipulation can be read without them.

\textbf{Sampling parameters.} None were set. Each call passes a model, the system prompt, the user prompt, and a \texttt{response\_\allowbreak format} of \texttt{json\_\allowbreak object}, so every generation ran at its provider's default temperature and top-p with JSON mode enforced. No seed was available on these routes.

\section{Clinical Anchors and the Frame Bound}
\label{app:anchors}

\begin{table*}[!b]
\caption{Published means across sampling frames, all expressed as PHQ-8, with the residual each implies. Rows 4 to 6 are PHQ-9 means converted by the NHANES item-9 offset matched to each anchor's own severity. NHANES administers the screener by interviewer, so those offsets are floors, each converted anchor an upper bound and each residual conservative. Both residual columns are computed on cisgender personas, the same frame as Table~\ref{tab:residuals}. The caseload row is reconstructed, not published, and its lower end is a defensible floor rather than an estimate. The Low SES column is not SES-matched; the matched comparison is +2.82 to +3.88. Rows 3 and 4 are printed to three decimals because their agreement at two is a coincidence of unrelated derivations.}
\label{tab:frame}
\small
\setlength{\tabcolsep}{3pt}
\begin{tabular}{llcc}
\toprule
Frame & Mean & Overall & Low SES \\
\midrule
General population, weighted (derived here) & 2.98 & +4.01 & +6.74 \\
Primary care, below severe on all three~\cite{lowe2008depression} & 3.60 & +3.39 & +6.12 \\
Primary care, full caseload (reconstructed)~\cite{lowe2008depression} & 4.583--5.641 & +1.35 to +2.41 & +4.08 to +5.14 \\
VA primary care, HIV-enriched cohort~\cite{stevens2020depression} & 5.644 & +1.35 & +4.08 \\
Internet panel, no clinical frame~\cite{perlis2025clinical} & 6.38 & +0.61 & +3.35 \\
Treatment entry, severity-gated~\cite{beck2011severity} & 11.95 & $-$4.96 & $-$2.22 \\
\bottomrule
\end{tabular}
\end{table*}

The prompt frames a screening encounter, and people completing a depression screen score higher than an unselected household sample. Some part of the residual in Table~\ref{tab:residuals} may be frame rather than miscalibration. Substituting a single clinic mean for a population parameter answers that badly, since clinic samples are selected on help-seeking, illness burden and insurance access, none of which this design implements. The literature supports a band, and Table~\ref{tab:frame} places the simulated means inside it.

Three anchors carry conditioning worth naming. L\"owe et al.~\cite{lowe2008depression} report 3.6 for the 1,759 patients, 84\% of their sample, scoring below 15 on the PHQ-8, the GAD-7 and the PHQ-15 alike; one of those three criteria is the dependent variable, so the figure conditions on the outcome being low. Beck et al.~\cite{beck2011severity} report 12.2 at antidepressant initiation among 771 patients who had to score 7 or higher to enter, conditioning on it being high. The VA figure of Stevens et al.~\cite{stevens2020depression} comes from a cohort recruited with roughly equal numbers of veterans with and without HIV and studied for comorbidity clustering, pooled over person-visits from 2003 to 2015. It conditions on comorbidity and bounds a general primary-care mean from above rather than estimating one.

The unconditioned quantity is the full primary-care caseload, which no source publishes. The decomposition runs the other way. Of the simulated cisgender overall mean's +4.01 points against the population, about 0.62 is the shift from the general population to a primary-care attendee below severe and about 1.5 is case mix. Against the reconstructed caseload interval of 4.583 to 5.641 the simulated 6.99 still leaves +1.35 to +2.41 that neither accounts for. Transferring the population socioeconomic gradient additively puts the low-income residual at +2.82 to +3.88, retaining 51\% to 71\% of the 5.48 points in Table~\ref{tab:residuals}. One entry sits close to the simulated mean: a US internet panel with no clinical framing~\cite{perlis2025clinical} converts to 6.38, against a simulated 6.99. That panel is a non-probability opt-in online sample rather than a probability sample of the population, and it carries the upward bias on depressive symptoms such samples are known for. Two things separate it from the simulations even at that distance. Dispersion is one: DeepSeek-V3 returns a standard deviation of 1.65 and GPT-4o-mini 2.11 against a population 3.94, and no human sample of any frame or mode compresses a 0--24 instrument that far. Shape is the other: the Kolmogorov--Smirnov distance against the population runs 0.43 to 0.73 for each model separately. A change of sampling frame moves location. It does not collapse scale.

The reconstruction rests on subgroup figures transcribed out of a closed-access table with no microdata behind them. No reader can recompute it. The primary comparison does not depend on it: Table~\ref{tab:residuals} is computed against microdata derived here and no clinical anchor enters it. An error there would change the size of the frame discount rather than the residual.

\textbf{Converting PHQ-9 anchors.} Most published US clinical means use the PHQ-9 on 0--27. The two instruments correlate at 0.996 to 0.998~\cite{wu2020equivalency} without being equal in level, differing by exactly one quantity, the mean of the ninth item, which we measure rather than assume. Under the NHANES survey weights it is 0.046 across adults and 0.056 among respondents scoring 5 to 8 on the PHQ-9, the band the screening anchors occupy, rising to 0.253 at treatment-entry severity. Where an anchor publishes its own tail fraction the offset is matched to that instead: the panel reports 26.4\% at or above 10, giving 0.123. Converting by these offsets moves an anchor by roughly a tenth of a point.

\textbf{Reconstructing the full caseload.} L\"owe et al.\ report by syndrome subgroup and the subgroups overlap, so no overall mean appears in their paper. Of the 332 patients outside the reference group, the 138 in the depression subgroup average 18.5, while the remaining 194 are elevated on anxiety or somatization and by construction score below 15 on the PHQ-8. The reported single-syndrome percentages reconstruct that 194 to within a tenth of a patient and the depression subgroup matches the reported 6.6\% prevalence exactly, so the decomposition is read as the authors built it. Bounding those 194 at zero is arithmetically valid and substantively impossible in a sample where 75\% of depressed patients carry a comorbid syndrome and the three scales intercorrelate at 0.64 to 0.75. Patients severe on a correlated syndrome do not average below patients severe on none. The reference mean of 3.6 is the conservative substitute and the working interval is 4.583 to 5.641.

\textbf{Matching the anchor to the persona.} The low-income persona reads \emph{``I make less than \$35,000 a year and I am on Medicaid''}, while the Low SES anchor of Table~\ref{tab:residuals} conditions on the poverty-income ratio below 1.30 and nothing else. Medicaid enrolment among adults is selected on disability and programme eligibility rather than income alone, and it carries its own severity. Recomputing the anchor on the same NHANES frame under four definitions gives 4.25 for the poverty ratio alone ($n = 10{,}663$), 5.45 for the poverty ratio crossed with Medicaid ($n = 2{,}545$), 4.02 for household income below \$35,000 ($n = 14{,}352$), and 5.52 for income below \$35,000 crossed with Medicaid ($n = 2{,}736$), the definition the persona states. The residual falls from $+5.48$ to $+4.21$, retaining 77\% of its magnitude. We report the poverty-ratio anchor as primary for continuity with the other rows of Table~\ref{tab:residuals}, all of which condition on demographics alone, and record the matched figure here as the conservative one.

\textbf{The socioeconomic transfer.} No SES-stratified clinical PHQ-8 anchor exists, so the low-income comparison transfers the population gradient onto the caseload band. The NHANES low-income anchor sits 1.26 points above the all-adults value, and we add that difference to each end of the band instead of scaling it, which assumes a clinical caseload's socioeconomic gradient matches the population's in absolute points. It is unmeasured. A steeper clinical gradient would raise the matched anchor and shrink the residual; a flatter one would do the reverse.

\section{Inference, Standard Errors and Null Constructions}
\label{app:inference}

\textbf{The unit and what pairing it does not license.} Pairing iteration $i$ of one condition with iteration $i$ of the other would impose a correspondence the design does not create. Condition contrasts are paired at the cell rather than at the iteration. Within-run stability statistics are the exception and are computed across the 30 iterations of a cell, where the iteration is the correct unit. Residuals are cell-mean averages minus ground-truth means, with 95\% intervals from the between-cell standard error. Treating generations as exchangeable understates standard errors by a factor of five to seven on the 60-generation model-by-cohort unit that pools both framings, and by four to five within a single condition; the algebraic ceiling on a 30-observation cell is $\sqrt{30} = 5.48$, so the larger figure belongs to the pooled unit only.

\textbf{The racial contrast test.} Section~\ref{sec:crossgroup} compares each racial contrast against the population value it should reproduce. A cell is a model crossed with a demographic cohort. A White cell and a Black cell sharing a model, condition, gender, socioeconomic level and relationship status differ in race alone. There are 96 such pairs per contrast, and the contrast is the mean paired difference, which removes the between-cohort variance that widens the marginal intervals. Two questions are separable and both are reported: whether the simulated contrast can be distinguished from zero, and whether it can be distinguished from the population contrast. The second carries the NHANES anchor error, entered as the root sum of squares of the two group design standard errors, available because the race groups are disjoint samples. Benjamini-Hochberg adjustment is applied across the three contrasts within each question. A stratum bootstrap resampling the twelve cohorts within each model, 10,000 draws, agrees on the Hispanic and Asian contrasts and disagrees on the Black one, where it returns $[-0.254, -0.011]$ and excludes zero. That bootstrap resamples cohorts inside each model and so holds the panel composition fixed, which suppresses exactly the variance the four models contribute by disagreeing in sign. The paired interval carries that variance and is the one we report. The gap between the two is the model-identity effect of Section~\ref{sec:modelfactor} appearing in an interval. The pairing key needs the condition in it: the raw socioeconomic field carries quoting variants from the two export code paths, which split three levels into six strings and would otherwise pair cells across conditions by accident.

\textbf{Anchor sampling error.} NHANES is a complex multistage sample. We compute its contribution by Taylor linearisation over the stratum and cluster identifiers in the demographic files, treating strata as SDMVSTRA and primary sampling units as SDMVPSU. Ignoring the design would understate these standard errors by a factor of 1.05 to 1.65 across the nine benchmarked anchors. The largest is 0.078 points, for the low-income anchor, and the widest resulting 95\% interval spans 0.31 points, an order of magnitude below the smallest effect under test. It is window-specific: on the 2021--2023 cycle the largest anchor error is 0.222, and each residual is compared against the error of its own window.

\textbf{Severity-controlled nulls.} Covariance divergence uses the Frobenius distance between PHQ-8 item correlation matrices, computed on severity-matched subsamples and referred to a null that applies the identical matched-subsample procedure to labels permuted within severity band. Null and statistic share both size and band composition. Dispersion differences across a design axis are tested twice, by permuting the axis labels within model and severity quintile, and by comparing cells matched on mean severity within model to a tolerance of 0.75 points. Both are needed because the simpler control, residualizing cell dispersion on cell mean with a pooled linear fit, is not adequate on a bounded scale: the relation is concave and model-specific, and socioeconomic status in this design is close to collinear with cell mean.

\textbf{The structural bootstrap.} Uncertainty in the band comparison comes from a cluster bootstrap that resamples design cells within each cohort and recomputes every matched distance per replicate. Because resampling with replacement leaves about 63\% of rows distinct and a Frobenius distance between two noisier estimates runs larger, the inflation falls hardest on the small contrasts and compresses every ratio toward one. Those intervals are reported as floors. The observed median sits at the 94th percentile of its own replicates and the observed extremum above all of them, and at the 2.5th percentile 88\% of the 49 pairings still order correctly.

\textbf{The bounded scale, and the detection floor on the interactions.} A 0--24 instrument compresses toward its centre, and Section~\ref{sec:overestimation} finds these models compressing ordinal responses toward the interior of every instrument they answer. The interaction result of Section~\ref{sec:structure} has to survive a change of scale. Refitting on $\mathrm{logit}(\text{total}/24)$, where a ceiling cannot manufacture curvature, holds the omnibus at $F = 4.07$ and moves the interaction variance from 1.31\% to 1.02\%. We report that as a single sensitivity and not a decomposition; Appendix~\ref{app:failed} records the two routes that would have apportioned it and why neither can. Attaching a detection floor requires a baseline the additive model actually fits, so we build one from the main-effects fit plus resampled residuals. On that baseline the nested $F$ test rejects at 8.3\% against a nominal 5\% and reaches 80\% power against an effect of 0.7 points, or 0.49\% of cell-mean variance against an observed 1.31\%. The surviving terms clear the floor by a factor of about two and a half.

\section{The Framing Contrast in Detail}
\label{app:framing}

Framing touches dispersion as well as level. Within-cell standard deviations run higher under clinical framing in all three SES bands, but the raw ratios overstate it. Clinical framing also raises the cell mean by 0.32 points, and cell dispersion tracks cell mean on this instrument at $\rho = 0.30$. Part of the gap is the severity shift reappearing as noise. Regressing dispersion on the cell mean and comparing the residuals leaves a gap of 0.076 points ($t = 3.66$, $p < 0.001$), roughly half the raw difference and sensitive to how the mean is controlled.

One reading of the whole contrast is that it measures the twelve days between collection sessions rather than the framing. The per-model pattern argues otherwise. A session or version confound has no reason to be model-specific, yet DeepSeek-V3 moves +0.74, GLM-4.7 and Gemini-3-Flash about +0.33, and GPT-4o-mini not at all. The endpoint carrying the acknowledged preview-route risk sits mid-pack while the largest mover sits on a pinned route, the ordering a session artefact would have to produce by coincidence. We read the heterogeneity as the load-bearing result and the pooled 0.32 as its least informative number.

\section{Where Regeneration Churn Lands}
\label{app:churn}

Table~\ref{tab:churn} shows where the crossings of Section~\ref{sec:fracture} land at fixed prompt on the left, and the same when framing is allowed to vary as well on the right. The cross-condition matrix has 1,724 cases moving from moderate under clinical framing to mild under narrative and 1,222 moving the other way, 55.8\% of its discordant pairs against the 56.0\% observed at fixed prompt. Adding framing variation redistributes almost none of the churn.

Instability also differs by instrument. Within a run, PHQ-8 and GAD-7 behave almost identically (35.4\% and 35.2\% flip probability under clinical framing, 32.2\% and 32.5\% under narrative), while AUDIT-C is markedly less stable at 48.1\% and 47.0\%, reflecting its three-item scale and narrow risk bands. PCL-5 is reported within-run only, since the two runs administered different forms.

\begin{table*}
\caption{Where regeneration churn lands. \textbf{Left:} severity-category agreement between two generations of the same cell at fixed prompt, clinical framing, over all 417,600 ordered within-cell iteration pairs (480 cells $\times$ 30 $\times$ 29; the 435 unordered pairs per cell are not independent observations, and no inference rests on this count); 64.60\% of pairs agree. \textbf{Right:} diagnostic category transitions between conditions, rows clinical and columns narrative, over 14,400 pairs; 9,121 of them (63.34\%) hold their category. Diagonal cells are agreements in both panels.}
\label{tab:churn}
\small
\setlength{\tabcolsep}{3pt}
\begin{minipage}[t]{0.48\linewidth}
\centering
\begin{tabular}{lrrrrr}
\toprule
 & None & Mild & Mod. & Mod-sev. & Sev. \\
\midrule
None (0--4) & 36,600 & 19,651 & 1,337 & 64 & 0 \\
Mild (5--9) & 19,651 & 149,736 & 41,411 & 1,919 & 27 \\
Moderate (10--14) & 1,337 & 41,411 & 63,074 & 9,263 & 45 \\
Mod-severe (15--19) & 64 & 1,919 & 9,263 & 20,340 & 198 \\
Severe (20--24) & 0 & 27 & 45 & 198 & 20 \\
\bottomrule
\end{tabular}
\end{minipage}\hfill
\begin{minipage}[t]{0.48\linewidth}
\centering
\begin{tabular}{lrrrrr}
\toprule
 & None & Mild & Mod. & Mod-sev. & Sev. \\
\midrule
None (0--4) & 1,453 & 484 & 49 & 2 & 0 \\
Mild (5--9) & 963 & 5,102 & 1,222 & 49 & 0 \\
Moderate (10--14) & 66 & 1,724 & 1,898 & 279 & 3 \\
Mod-severe (15--19) & 4 & 89 & 332 & 668 & 3 \\
Severe (20--24) & 0 & 1 & 1 & 8 & 0 \\
\bottomrule
\end{tabular}
\end{minipage}
\end{table*}

\section{Per-Contrast Covariance Results}
\label{app:contrasts}

Table~\ref{tab:contrasts} lists every contrast behind Sections~\ref{sec:covariance} and~\ref{sec:structure}. Raw distances are computed on full cohorts; matched distances on severity-matched subsamples against a null built by applying the same matched-subsample procedure to labels permuted within severity band. The last three columns repeat the matched statistic with every design cell centred on its own item means, each against its own recomputed null, since the centring changes the noise floor and a centred statistic read against an uncentred null would report that floor as structure.

\begin{table*}[b]
\caption{Every cohort contrast in the design: raw Frobenius distance, the severity-matched statistic, the 95th percentile of its size-matched permutation null, and the permutation p-value, then the same three quantities with between-cell mean structure removed. All fourteen exceed their nulls on both statistics. The centred column moves in both directions: gender contrasts fall, socioeconomic and racial ones rise, and the rise is the noise a mean estimated from 30 draws adds to every correlation. Excess over the matched null is the quantity to compare, and it is measured against the null mean rather than the p95 printed here. The excess figures in Section~\ref{sec:structure} are larger than this table's columns differenced. Matched and centred figures come from independent runs of the same procedure, so the two matched columns differ by up to 0.03 through resampling alone. Multiracial against White is the one contrast whose matched value exceeds its raw value (0.19 to 0.203), the size-dependence of the statistic showing through.}
\label{tab:contrasts}
\small
\setlength{\tabcolsep}{5pt}
\begin{tabular}{lrrrrrrr}
\toprule
 & \multicolumn{4}{c}{Pooled cohort} & \multicolumn{3}{c}{Cell-centred} \\
\cmidrule(lr){2-5}\cmidrule(lr){6-8}
Contrast & Raw & Matched & Null p95 & $p$ & Matched & Null p95 & $p$ \\
\midrule
Low vs High & 1.23 & 0.801 & 0.193 & .001 & 0.812 & 0.198 & .002 \\
Cis man vs Trans woman & 1.14 & 0.539 & 0.178 & .001 & 0.463 & 0.198 & .002 \\
Low vs Middle & 0.88 & 0.536 & 0.162 & .001 & 0.434 & 0.175 & .002 \\
Cis man vs Trans man & 1.06 & 0.470 & 0.171 & .001 & 0.383 & 0.191 & .002 \\
Cis woman vs Trans woman & 0.89 & 0.464 & 0.168 & .001 & 0.296 & 0.190 & .002 \\
Cis woman vs Trans man & 0.82 & 0.409 & 0.161 & .001 & 0.251 & 0.184 & .002 \\
Middle vs High & 0.63 & 0.367 & 0.137 & .001 & 0.542 & 0.155 & .002 \\
Cis man vs Cis woman & 0.32 & 0.210 & 0.146 & .001 & 0.227 & 0.172 & .002 \\
Black vs White & 0.24 & 0.205 & 0.144 & .001 & 0.195 & 0.188 & .035 \\
Multiracial vs White & 0.19 & 0.203 & 0.142 & .001 & 0.241 & 0.188 & .002 \\
Hispanic vs White & 0.23 & 0.198 & 0.141 & .001 & 0.236 & 0.186 & .002 \\
Asian vs White & 0.37 & 0.170 & 0.152 & .010 & 0.231 & 0.194 & .002 \\
Trans woman vs Trans man & 0.17 & 0.157 & 0.152 & .027 & 0.170 & 0.166 & .035 \\
Single vs Married & 0.25 & 0.142 & 0.103 & .001 & 0.136 & 0.123 & .007 \\
\bottomrule
\end{tabular}
\end{table*}

\section{Statistical Test Ledger}
\label{app:ledger}

Table~\ref{tab:ledger} summarizes the 34-test family with Benjamini-Hochberg adjustment; the machine-readable ledger with exact p-values, seeds and per-test inputs ships with the release. Thirty-three of the 34 clear $q < 0.05$. The exception is T29, the GPT-4o-mini condition contrast, the one comparison in the paper that does not reach significance once the design cell is used as the unit. That the rest clear is expected at this sample size, and is why Section~\ref{sec:covariance} interprets the covariance results against severity-matched permutation nulls and reports magnitudes rather than resting on these p-values.

\begin{table}
\caption{Test family summary, BH-adjusted across all 34 tests. All but T29 clear $q < 0.05$. Every test uses the design cell as its unit.}
\label{tab:ledger}
\small
\setlength{\tabcolsep}{3.5pt}
\begin{tabular}{lll}
\toprule
Tests & Family & Type \\
\midrule
T01--T09 & Residuals vs ground truth & one-sample t \\
T10, T25--T26 & Condition contrast, three scales & paired t \\
T27--T30 & Condition contrast per model\textsuperscript{a} & paired t \\
T11 & Cross-condition flip & Kruskal-Wallis\textsuperscript{c} \\
T12 & Gateway violations & Kruskal-Wallis\textsuperscript{c} \\
T13--T14 & Trans-cis elevation & Welch t \\
T15--T24 & Covariance, race and gender\textsuperscript{b} & permutation \\
T31--T33 & Covariance, SES & permutation \\
T34 & Covariance, relationship & permutation \\
\bottomrule
\end{tabular}

\smallskip
\raggedright\footnotesize
\textsuperscript{a}\,GPT-4o-mini's contrast (T29) is the family's only null: $t = -1.48$, $q = .14$.\\
\textsuperscript{b}\,Ledger entries T15--T34 are raw-distance tests; the severity-matched comparisons of Section~\ref{sec:covariance} carry their own size-matched nulls.\\
\textsuperscript{c}\,Computed on the 480 per-cell proportions, not on generations, for the reason given in Section~\ref{sec:stats}.
\end{table}

\section{Model Behavioural Profiles}
\label{app:profiles}

\begin{figure}
\includegraphics[width=\linewidth]{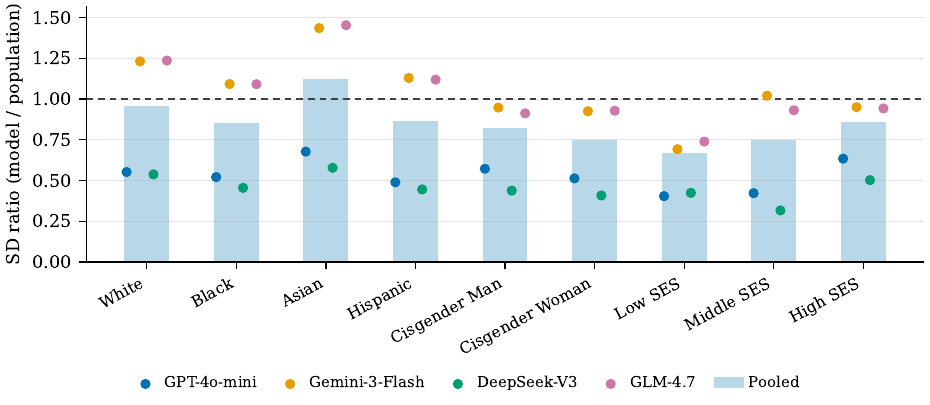}
\caption{Output-to-population SD ratios. The population SD is a common scale here and not a fidelity target, for the reason Section~\ref{sec:estimand} gives. Bars: four models pooled. Markers: individual models. Values below the dashed line indicate narrower-than-population output distributions.}
\label{fig:variance}
\end{figure}

\textbf{Dispersion, descriptively.} Output-to-population SD ratios are descriptive comparisons and not fidelity measures (Section~\ref{sec:estimand}). Across groups the low-SES cell shows the single narrowest output distribution relative to its population counterpart (ratio 0.67, with SES cells spanning 0.67 to 0.86) and Asian personas the widest (1.12, Figure~\ref{fig:variance}). Across models the ratios split cleanly: DeepSeek-V3 (0.46) and GPT-4o-mini (0.53) produce narrow output distributions, while GLM-4.7 (1.04) and Gemini-3-Flash (1.05) sit at population-level dispersion. We read that split as reasoning mode until the call ledger settled it. Only GLM-4.7 bills reasoning tokens, a mean of 1,400 per call against none for the other three, and Gemini-3-Flash reaches the same dispersion without spending any. Whatever separates the two pairs, a reasoning pass is not it. With four models divided two and two, any single ranking separates the groups cleanly with probability one in three. Across the three model-level measures we report, one clean separation is exactly what chance predicts, and the other two do interleave. The magnitude structure is harder to get by chance, the gap between groups being 0.51 against a wider within-group spread of 0.07. The design cannot separate any of this from training or alignment differences, since reasoning status now falls one against three and every other model-level property falls the same way.

\begin{figure}[!t]
\includegraphics[width=0.8\linewidth]{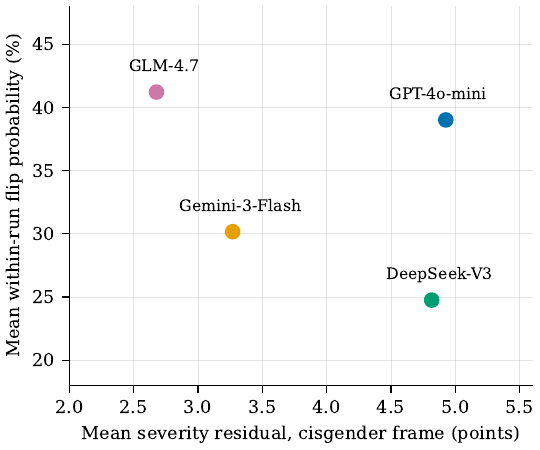}
\caption{Model behavioural profiles: mean severity residual (cisgender frame) against mean within-run flip probability. No model dominates, and the four do not form a frontier.}
\label{fig:profiles}
\end{figure}

\textbf{Calibration against stability.} The per-model receipts yield four distinct profiles (Figure~\ref{fig:profiles}). GLM-4.7 carries the smallest mean calibration residual (2.68 points, cisgender frame) but the worst stability (41\% mean within-run flip probability) and the worst coherence (6.24\% gateway violations). DeepSeek-V3 inverts the profile: most stable (25\%) and most coherent (0.35\%), but among the most severity-inflated (4.82). GPT-4o-mini pairs the largest residual (4.93) with poor stability (39\%), and Gemini-3-Flash comes closest to balancing the three (3.27 points, 30\%, 0.55\%). These four do not form an accuracy-stability frontier: GPT-4o-mini is worse than Gemini-3-Flash on both axes at once. Model selection for simulation work is multi-objective, no model in this set dominates, and a pipeline that selects on calibration alone will import an instability problem it never measured.

\section{Distribution Shape and Alternative Comparators}
\label{app:shape}

\begin{figure}
\includegraphics[width=\linewidth]{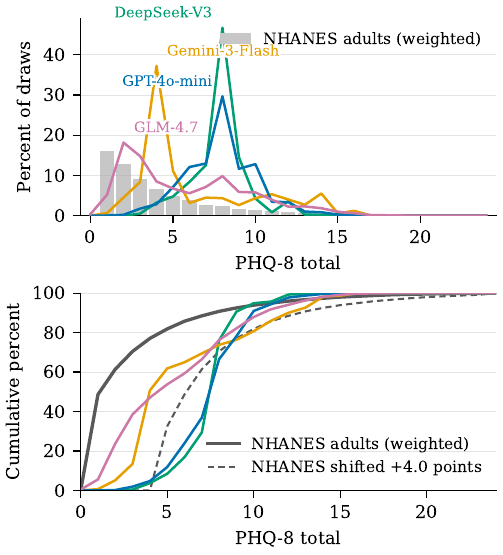}
\caption{Simulated and population PHQ-8 distributions, cisgender personas. Top: response distributions, NHANES weighted in grey, each model direct-labelled. Bottom: cumulative distributions, with the population distribution also shown shifted upward by the pooled mean residual (dashed). The dashed curve shows where a pure level error would put the population; the vertical distance from it to each model curve is illustrative and we attach no decomposition to it.}
\label{fig:distribution}
\end{figure}

Figure~\ref{fig:distribution} shows the distributions behind Section~\ref{sec:overestimation}. The concentration is far tighter in two of the four models than in the pool: 86.9\% of DeepSeek-V3 draws and 73.4\% of GPT-4o-mini draws fall in the single five-point 5--9 window, at standard deviations of 1.65 and 2.11 against a population 3.94. Sixteen of the 28,800 generations reach the severe range and every one is a transgender persona, ten under clinical framing and six under narrative, with fourteen of the sixteen coming from GLM-4.7. A pure response-format effect should suppress the top category uniformly, since the format does not know which persona it is answering for. Those sixteen are weak evidence that persona content can overcome format-level suppression. Sixteen cases is a thin base and the observation stays qualitative.

\section{Factor Structure Under Two Estimators}
\label{app:factor}

Section~\ref{sec:covariance} argues that a factor solution fitted to pooled cohorts will describe none of them, and that consequence is testable. It is also where this paper and Villarreal-Zegarra and Bellido-Boza~\cite{villarrealzegarra2026synthetic} might be reconciled.

Fitting one common single-factor specification within each of the twelve marginal cohorts and within the NHANES adult sample, the population sample reaches SRMR 0.042 and no simulated cohort does, running 0.093 to 0.125 pooled over models. Inside a single model, nine of twelve cohorts clear 0.08 in Gemini-3-Flash and nine in GLM-4.7, while none clear it in DeepSeek-V3 or GPT-4o-mini. Agreement between the two studies is therefore model-dependent.

Maximum likelihood on Pearson correlations of four-category items is not the estimator a psychometric study would choose, and we expected a weighted least squares variant to fit all of these better. Refitting on polychoric correlations by diagonally weighted least squares, which is the estimation half of WLSMV, shows that it does not. The population fit improves slightly, SRMR 0.0421 to 0.0411, while every simulated cohort gets worse: the twelve pooled cohorts move from 0.093--0.125 to 0.108--0.150, and the within-model counts fall from nine of twelve to six in Gemini-3-Flash and from nine to four in GLM-4.7. De-attenuating the correlations does not close the gap, because what separates these matrices from the population's is not attenuation. Two limits on that refit: SRMR is a residual and not a test statistic, so the mean- and variance-adjusted chi-square that completes WLSMV never enters, and the weights hold each item's thresholds at their marginal estimates.

\section{Diagnostics That Failed}
\label{app:failed}

We tried two diagnostics that did not work, and record them here.

\textbf{A single number for the distributional gap.} The natural candidate, the share of the Kolmogorov--Smirnov distance that a location shift removes, is unusable on this instrument. Both distributions sit on the integers 0 to 24 and the statistic turns on the fractional part of the shift. On data constructed to be a pure location shift with no shape difference whatever, rounded back onto the same support, it reports between 55\% and 99\% of the distance closed depending on where the empirical mean falls between two lattice points.

\textbf{Apportioning the interaction signal between compression and compounding.} We first pushed the additive fit on the logit scale back through the inverse link and correlated the departure pattern that transformation induces against the observed departures, giving $r = 0.41$. We withdraw it. On a balanced factorial the second-order term of the inverse link is a bilinear form in the marginal offsets, and over these 120 cohorts that induced pattern correlates at $r = 0.86$ with a pure bilinear interaction kernel built from the same offsets and carrying no information about any bound. The regressor cannot separate a compressed scale from compounding axes, so its correlation with the observed departures apportions nothing. The diagnostic it replaced failed the same way for a different reason: it regressed the fitted values of an additive model on its own residuals, which returns zero by construction.

\FloatBarrier

\section{Decoding Control}
\label{app:decoding}

The regeneration result of Section~\ref{sec:fracture} was measured at each provider's default decoding, because no sampling parameter was ever set. This control measures what setting temperature to 0 would change.

The design was fixed before any call was made and is released with the data. Four endpoints, twelve cohorts, thirty draws and two decoding arms give 2,880 generations under clinical framing. Cohorts were drawn one per stratum after ranking all 120 by their observed clinical mean PHQ-8 under a recorded seed. Stratifying on the released means spans the severity range without selecting on the outcome the experiment measures. Both arms were collected in the same session, which makes the contrast internal to a single window.

Category flipping falls on three endpoints and holds on the fourth. Pooled over models it runs 31.71\% at provider defaults and 13.13\% at temperature 0. Table~\ref{tab:decoding} gives the per-model figures alongside the determinism diagnostic that explains them.

\begin{table*}[!t]
\centering
\caption{Decoding control, 2,880 pre-registered generations over four endpoints, twelve cohorts and thirty draws per arm. Upper panel: within-cell severity-category flip probability under each decoding arm, with the two diagnostics that separate a steadied judgement from a silenced sampler. Change carries a 95\% interval from a cluster bootstrap resampling the twelve cohorts, both arms travelling together. Modal share is the mean share of a cell taken by its most common total at temperature 0; one-valued cells are those returning a single total across all thirty draws. Lower panel: the default arm against the December corpus on the same design, eight months earlier.}
\label{tab:decoding}
\small
\begin{tabular}{lrrrrr}
\toprule
& \multicolumn{3}{c}{Category flipping (\%)} & \multicolumn{2}{c}{Temperature 0 output (\%)} \\
\cmidrule(lr){2-4}\cmidrule(lr){5-6}
Model & Default & Temp 0 & Change & Modal share & One-valued \\
\midrule
GPT-4o-mini    & 37.32 &  1.55 & $-35.8$ [$-41.8$, $-29.1$] & 90.6 & 58.3 \\
Gemini-3-Flash & 23.47 &  4.14 & $-19.3$ [$-29.9$, $-10.2$] & 89.7 & 75.0 \\
DeepSeek-V3    & 24.64 &  9.06 & $-15.6$ [$-23.9$, $-8.1$]  & 63.6 &  0.0 \\
GLM-4.7        & 41.42 & 37.76 & $-3.7$ [$-9.8$, $+2.0$]    & 34.4 &  0.0 \\
\midrule
Pooled         & 31.71 & 13.13 & $-18.6$                    &      &      \\
\bottomrule
\end{tabular}

\vspace{1em}

\begin{tabular}{lrrrrrr}
\toprule
& \multicolumn{3}{c}{Cohort mean PHQ-8} & \multicolumn{3}{c}{Category flipping (\%)} \\
\cmidrule(lr){2-4}\cmidrule(lr){5-7}
Model & Dec & Aug & Change & Dec & Aug & Change \\
\midrule
GPT-4o-mini    & 8.32 & 8.14 & $-0.19$ & 37.26 & 37.32 & $+0.06$ \\
Gemini-3-Flash & 9.07 & 9.02 & $-0.05$ & 29.16 & 23.47 & $-5.69$ \\
DeepSeek-V3    & 8.91 & 8.56 & $-0.35$ & 25.75 & 24.64 & $-1.11$ \\
GLM-4.7        & 8.27 & 8.74 & $+0.47$ & 46.53 & 41.42 & $-5.11$ \\
\bottomrule
\end{tabular}
\end{table*}

The mechanism differs by endpoint, and the pooled figure describes none of the four. Gemini-3-Flash returns one repeated total in three quarters of its cells at temperature 0 and GPT-4o-mini in well over half, their distinct item vectors per thirty draws collapsing to 1.5 and 1.9. Neither model steadied a judgement. Both stopped producing evidence about one. DeepSeek-V3 narrows genuinely, with no cell collapsing to a single value, while GLM-4.7 barely moves at all. Bootstrapping the twelve cohorts puts GLM-4.7's change at $-3.66$ with a 95\% interval of $-9.75$ to $+1.95$, so on that endpoint temperature 0 is not distinguishable from leaving the decoding alone. The interval separates GPT-4o-mini from DeepSeek-V3 and GLM-4.7, and Gemini-3-Flash from GLM-4.7; the remaining pairs overlap and the design does not rank them. We take that to mean a decoding instruction is a request and not a guarantee. A deployment cannot verify compliance without measuring it.

Greedy decoding does not resolve the cases sitting near the treatment boundary. In 15 of the 48 model-cohort cells the default arm was itself split between 20\% and 80\% across PHQ-8 = 10. In those cells temperature 0 reports one side carrying no indication that the case was borderline. The temperature-0 answer is also not the centre of the default distribution. Its median percentile within that distribution is 43.3, it sits 21.2 percentile points off the midpoint on average, and it falls outside the interquartile range in 35.4\% of cells. Instability at temperature 0 still covers 18 of 48 cells, 8 of them straddling PHQ-8 = 10, with GLM-4.7 accounting for 12.

The structural results reported in the body do not depend on the decoding arm. No cell moves by 5 points or more between arms. The temperature-0 category matches the default modal category in 89.6\% of cells and falls on the default majority side of PHQ-8 = 10 in 93.8\%. Cohort ordering agrees between arms at Spearman 0.850 to 0.979 within model. Cross-model ordering agreement runs 0.925 at defaults against 0.891 at temperature 0. The remedy therefore degrades agreement between systems slightly instead of improving it.

The default arm also re-measures the December corpus on the same design eight months later, and the lower panel of Table~\ref{tab:decoding} carries that comparison. Across the twelve cohorts the means shift by 0.23 points on average, with no cohort moving more than 0.50. A rank correlation is also available and reads 0.986, but the cohorts were drawn one per stratum after ranking on this same variable, so they span the severity range by construction and the rank agreement is partly an artefact of that selection. The mean absolute shift carries the result. The generation layer used here was rebuilt from Appendix~\ref{app:prompts}, the original not having been preserved. That agreement therefore bears on whether the appendix suffices to regenerate cohort-level structure. It is not a reproduction of the corpus and the pre-registration forbids reporting one.

Four limits bound what this control shows. Two of the four routes are undated substitutes for December snapshots since retired, and a per-model match indicates a behaviourally close build instead of the same build. GLM-4.7 was additionally pinned to one upstream provider to hold quantization constant across arms. That provider served none of the December collection, and the GLM replication row therefore spans a provider change as well as an endpoint change. The sample is twelve cohorts of 120 under clinical framing alone. The ledger closes to the row under a content-independent rule. Of 3,520 raw lines, 6 were API failures and 2 were unparsed. A further 632 duplicate draw keys were dropped after concurrent writers duplicated part of the collection, leaving 2,880 generations at exactly thirty per cell.

\FloatBarrier

\end{document}